\documentclass[manuscript]{aastex62}

\usepackage{amsmath, amssymb}
\usepackage{booktabs}

\turnoffediting

\newcommand{\ahzin}{a_{\text{IHZ}}}

\newcommand{\ahzout}{a_{\text{OHZ}}}

\newcommand{\acrit}{a_{\text{crit}}}

\newcommand{\eq}[1]{eq.~(\ref{#1})}

\received{XXX, 2019}
\revised{YYY, 2020}
\accepted{ZZZ, 2020}

\submitjournal{AJ}

\shorttitle{Shadows of the Giants}
\shortauthors{Bazs{\'o} et al.}

\begin{document}

\title{Fear the Shadows of the Giants: On Secular Perturbations in Circumstellar Habitable Zones of Double Stars}

\correspondingauthor{{\'A}kos Bazs{\'o}}
\email{akos.bazso@univie.ac.at}

\author{{\'A}kos Bazs{\'o}}
\affil{%
    Department of Astrophysics \\%
    University of Vienna \\%
    T{\"u}rkenschanzstrasse 17 \\%
    A-1180 Vienna, Austria%
}

\author{Elke Pilat-Lohinger}
\affil{%
    Department of Astrophysics \\%
    University of Vienna \\%
    T{\"u}rkenschanzstrasse 17 \\%
    A-1180 Vienna, Austria%
}





\begin{abstract}
After the detection of extrasolar planets in binary and multiple star systems questions arose about their dynamics and habitability.
In this study we investigate a five-dimensional parameter space for wide binary stars with a massive planet beyond the habitable zone (HZ).
Our aim is to reveal those orbital and physical parameter combinations that expose bodies in the primary star's HZ to secular perturbations.
Building upon an established semi-analytical model, we combine two separate analytical models into a new one to treat the restricted four-body problem.
We apply this new Combined Analytical Model (CAM) to different synthetic systems and systematically map the occurrence of secular resonances (SR).
These maps are then visualized as two-dimensional sections of the parameter space.
The CAM model has a median error below 3~\% relative to numerical reference simulations.
We also derive a simplified CAM that performs well for hierarchical systems with moderate to large separations between the bodies.
Our results show that SR appear in the HZ even for large secondary star distances (up to $> 1000$~au) if either (i) the planet's distance is larger than Jupiter's, or (ii) its mass is about Saturn's or lower.
Changes in the secondary star's eccentricity by stellar flybys or galactic tides can push a formerly dynamically quiet HZ to a high-eccentricity state.
Based on these results we provide the easy-to-use online tool `SHaDoS' that evaluates the CAM for a given user input and traces the locations of SR in the HZ in two-dimensional parameter space plots.
\end{abstract}

\keywords{%
    binaries: general;
    celestial mechanics: restricted four-body problem;
    habitability;
    methods: analytical;
    planets and satellites: dynamical evolution and stability;
    planets and satellites: terrestrial planets
}



\section{Introduction} \label{sec:intro}


Observational surveys revealed that a considerable fraction of stars in the solar neighbourhood are members of binary and multiple star systems.
In this context, \citet{Duq1991} and \citet{Rag2010} established that in the solar neighbourhood \edit1{up to  distances of $d \le 25$~pc} about $40-45$~\% of all Sun-like stars \edit1{with spectral types F6--K3} are members of binary and multiple star systems, independent of whether or not they are hosting extrasolar planets (or short exoplanets).
\citet{Tok2014} found for a sample of about 4800 F- and G-type main-sequence stars within 67~pc of the Sun that 33~\% of the targets belong to binary star systems.
Further details on the frequency and main characteristics of multiple star systems can be found in the review of \citet{Duc2013}.


The detection of exoplanets also in binary and multi-stellar systems showed that planetary companions are not restricted to single stars.
However, it is uncertain how favourable stellar multiplicity really is for the presence of exoplanets.
Based on the discovered exoplanet population \citet{Rag2006}, \citet{Des2007}, \citet{Mug2009}, and \citet{Roe2012} estimated the stellar multiplicity.
These studies yield that about $10-15$~\% of the detected exoplanets reside in binaries, while \edit1{only} $\sim 2$~\% of all \edit1{known} exoplanet host stars are members of multiple star systems with three or more components.
The {\it Catalogue of Exoplanets in Binary and Multiple Star Systems}\footnote{https://www.univie.ac.at/adg/schwarz/multiple.html} of \citet{Sch2016} provides an up-to-date compilation of all binary and multiple star systems hosting exoplanets.


Although many exoplanets have been found in both binary and multiple star systems, it is unclear whether or not these environments are more hostile than single star systems for the presence of planets.
There are different views on the efficiency of planet formation in binary stars and the impact of the stellar companion's perturbations.
On the one hand, simulations of protoplanetary disks were able to form both terrestrial and giant planets even in tight binary systems, see e.g. \citet{Boss2006, Qui2007, Raf2013, Bro2015, Jang2015}.
These studies relate to rather tight systems with stellar separations of $< 100$~astronomical units (au), such as the poster cases $\gamma$~Cephei and HD~41004 with $\sim 20$~au.
On the other hand there are indications that the occurrence rate of circumstellar planets is strongly influenced by close binary companions \citep[see][]{Kley2010, Mul2012, Wang2014, Kra2016}.
For more comprehensive reviews on planet formation in binary star systems refer to \citet{Zhou2012}, \citet{The2015}, and \citet{Pil2019}.


Apart from the circumstances of planet formation in binaries also the long-term stability of planetary systems is an issue.
In this study we focus exclusively on circumstellar (S-type) planetary systems of binary stars \citep{Dvo1982}.
Different kinds of resonances play a key role for the orbital evolution of the planets, \edit1{e.g. mean motion and secular resonances}.

Mean motion resonances (MMR) \edit1{involve combinations of orbital frequencies and their overlap has already been identified to be crucial in triggering chaos} \citep{Mud2006, Had2018}.
The occurrence of circumstellar planets depends sensitively on the interactions with the outer stellar companion and on its orbital parameters \citep{Rabl1988, Hol1999, Pil2002}.
\edit1{%
All these studies have focused on the stability of prograde planets though, but retrograde resonances offer a stabilizing mechanism beyond the established stability borders for the prograde case.
\citet{Mor2012} and \citet{Mor2013} studied retrograde resonances for binary stars in the context of the circular spatial restricted three-body problem.
These works were based on the study of \citet{Gay2008} who demonstrated that retrograde MMR can stabilize compact multi-planetary systems.
In the former studies of binary systems the (retrograde) resonant terms were identified in the expansion of the disturbing function and two main aspects of the stabilizing mechanism were discussed: first, the eccentricity forcing for a prograde planet is larger than for a retrograde one (i.e. different exponents for the eccentricity in the leading term of the expansion), and second the overlap among nearby resonances occurs at larger perturber masses for retrograde planets.
}

Secular interactions \edit1{are related to apsidal and nodal precession frequencies and constitute another way how} the stellar companion can influence circumstellar planetary systems.
\edit1{%
Similar to MMR, SR affect not only prograde but retrograde orbits as well, and can even induce a switch between these two types of orbital motion.
\citet{Naoz2013} investigated the secular dynamics of hierarchical three-body systems and derived an octupole order secular approximation to the equations of motion.
From their model they found a qualitatively different dynamical evolution for an inner body that can become highly eccentric and would then oscillate between prograde and retrograde motion when it is perturbed by an eccentric outer body.
\citet{Li2017} applied a semi-analytical model to investigate SR for prograde and retrograde orbits in a restricted four-body problem.
They identified the critical angles of the strongest SR that are formed by linear combinations of apsidal and nodal differences of the two adjacent orbits.
}
In general, even initially circular and coplanar planetary orbits can be strongly altered by secular perturbations from a massive, eccentric and inclined outer companion.
\citet{Tak2008} investigated the angular momentum transfer from a stellar companion to weakly or rigidly coupled planetary systems.
They identified that this cascaded angular momentum transfer through the system leads to a significant eccentricity excitation for the innermost planet for a certain range of outer planet parameters.
In a similar study, \citet{Pu2018} found that eccentricities and mutual inclinations of weakly coupled multi-planet systems are more effectively excited by external companions than for their strongly coupled counterparts (i.e. compact planetary systems).

Still another type of resonant forcing exerted by the binary companion involves the evection resonance \citep{Fro2010, Tou2015} that combines the orbital precession frequency of an inner planetary orbit with the orbital motion of the outer secondary star (or another perturber).
This process can lead to large amplitude oscillations of the planetary eccentricities and mutual inclinations, or even to the disruption of the planetary system.
Recently, \citet{Mar2016} demonstrated that the outer stability border of planetary orbits in multi-planet systems of binaries is destabilized by overlap of MMR and SR (both linear and non-linear ones).


In spite of all these different mechanisms to disrupt planetary systems it is not hopeless to find terrestrial planets in binaries, preferably even in the habitable zone (HZ).
\citet{Funk2015} examined four different tight binary systems with a known massive planet to explore their potential of being able to host additional habitable terrestrial planets.
They concluded that two systems are promising candidates and permit dynamically stable orbits in the HZ.
However, the identification of suitable systems requires a thorough knowledge of the locations of resonances.
In case of MMR this is trivial, but it is more difficult to find the locations of SR.
To tackle this task, \citet{Pil2016} and \citet{Baz2017} \edit1{described} a semi-analytical method that simplifies matters.
That method determines the orbital precession frequencies for the perturbers and allows to estimate the positions of linear SR for test particles.
However, the method is valid for the system as it is, i.e. for the instantaneous orbital parameters of the perturbers; if the parameters change with time or by improved observations, then the analysis would need to be repeated.

In this work we aim at a more global description of the dynamics in a binary star system with a giant planet.
Our goal is to reveal parameter combinations that expose bodies (terrestrial planets) in the primary star's HZ to secular perturbations.
These parameters cover all relevant observables, in particular masses, orbital distances, and eccentricities, such that observational uncertainties are implicitly included into the analysis.
We investigate this multi-dimensional parameter space by means of a combination of analytical models.
This allows to quickly identify \edit1{the locations of} SR and to \edit1{indicate whether to} safely exclude systems with a-priori perturbed HZs \edit1{or to perform} a subsequent detailed stability analysis \edit1{with complementary tools}.
A possible application of the method is for newly discovered exoplanetary systems of binary stars, where it would help to exclude those systems that cannot host planets in the HZ in low eccentricity orbits.
This, in turn, renders the method especially valuable as a guide to select such binary star systems for in-depth observations that would allow planets in the HZ from a dynamical point of view.


The outline of the paper is as follows.
We describe the basic dynamical model along with the analytical methods employed for the parameter search in section~\ref{sec:method}.
Subsequently, in section~\ref{sec:result}, we first investigate the performance of the analytical models relative to reference data from numerical simulations, and then present the results on the SR location depending on the system parameters.
Section~\ref{sec:discuss} contains a discussion of the results and how they compare to other studies.
Finally, section~\ref{sec:summary} highlights the main results.
Additionally, in Appendix~\ref{sec:app:tool} we introduce an online tool that is based on the models from section~\ref{sec:method} and can be used to create plots similar to those in this article.




\section{Theory and models} \label{sec:method}

Our main goal is to identify those combinations of physical and orbital parameters for binary star systems with a circumstellar giant planet that lead to secular perturbations inside the host star's habitable zone.
First, we provide the HZ model and state the HZ location and its extent in section \ref{sec:method:hz}.
Next, we introduce the dynamical model for the secular evolution of the system in section \ref{sec:method:dynmod}, and define the main relation for the secular frequencies.
\edit1{On the basis of this relation we show how to use the combined analytical model (CAM) that consists of two consecutive steps, each of them given by a simple analytical expression, that are described in sections \ref{sec:method:andegg} and \ref{sec:method:laplag}.}
Finally, we demonstrate some of the properties of the CAM and derive a simplified form for it.
\edit1{Table~\ref{tab:notation} contains a list of the variable names used in this work, together with a short description for a quick overview.}

\begin{deluxetable}{ll}
    \tablecaption{Glossary of important variables used in this work.\label{tab:notation}}
    \tablehead{\colhead{Variable} & \colhead{Description}}
    \startdata
    $a_{B}$         & semi-major axis of secondary star (component B) \\
    $a_{GP}$        & semi-major axis of giant planet \\
    $\ahzin$        & inner border of host star's habitable zone \\
    $\ahzout$       & outer border of host star's habitable zone \\
    $a_{TP}$        & semi-major axis of test planet \\
    $\alpha$        & semi-major axis ratio, e.g. $a_{GP} / a_{B}$ \\
    $b_{3/2}^{(1)}$ & Laplace coefficient \\
    $\gamma$        & mass ratio, e.g. $m_{B} / m_{A}$ \\
    $\delta_{g}$    & empirical correction term for the secular frequency $g$, see \eq{eq:andeggcorr} \\
    $e_{B}$         & eccentricity of secondary star (component B) \\
    $e_{GP}$        & eccentricity of giant planet \\
    $g_{GP}$        & secular frequency of giant planet \\
    $g_{TP}$        & secular frequency of test planet \\
    $m_{A}$         & mass of primary star (component A) \\
    $m_{B}$         & mass of secondary star (component B) \\
    $m_{GP}$        & mass of giant planet \\
    $m_{TP}$        & mass of test planet, equals zero by definition \\
    $n_{GP}$        & mean motion of giant planet \\
    $n_{TP}$        & mean motion of test planet \\
    \enddata
\end{deluxetable}


\subsection{Habitable zone} \label{sec:method:hz}

There exist various models for calculating HZs for binary stars and in particular circumstellar HZs, e.g. \citet{Eggl2012}, \citet{Kal2013}, \citet{Cun2014}, and its update in \citet{Wang2019}.
A common point of all these models is that they extend the classical concept of habitable zones of single stars, see e.g. \citet{Kas1993} or \citet{Kop2014}, by taking into account multiple sources of radiation, and to some extent also the dynamical evolution of objects.

In case of circumstellar planets, the HZ location still primarily depends on the luminosity (mass) and effective temperature of the host star.
Thus, from a practical point of view, it is permissible to approximate the binary star HZ as that of a single star HZ (SSHZ), because in the following we will typically deal with rather large stellar separations (on the order of some hundred au).
In such cases the secondary's contribution is practically negligible; for stellar separations $> 20$~au the binary star HZ borders typically shift by $< 0.1$~au relative to the SSHZ.

For our purposes we fix the host star's spectral type\footnote{Without loss of generality; this choice only affects the HZ limits that can be easily calculated for other spectral types (masses).} to G2~V, which gives a solar analogue star with mass $1 \; M_{\odot}$, luminosity $1 \; L_{\odot}$, and effective temperature 5780~K.
With this set of parameters we obtain a HZ that ranges from the inner HZ border (IHZ), $\ahzin = 0.95$~au, to the outer HZ border (OHZ), $\ahzout = 1.68$~au, according to the model of \citet{Kop2014}.


\subsection{Dynamical model} \label{sec:method:dynmod}

\begin{figure}
    \centering
    \includegraphics[width=0.5\textwidth]{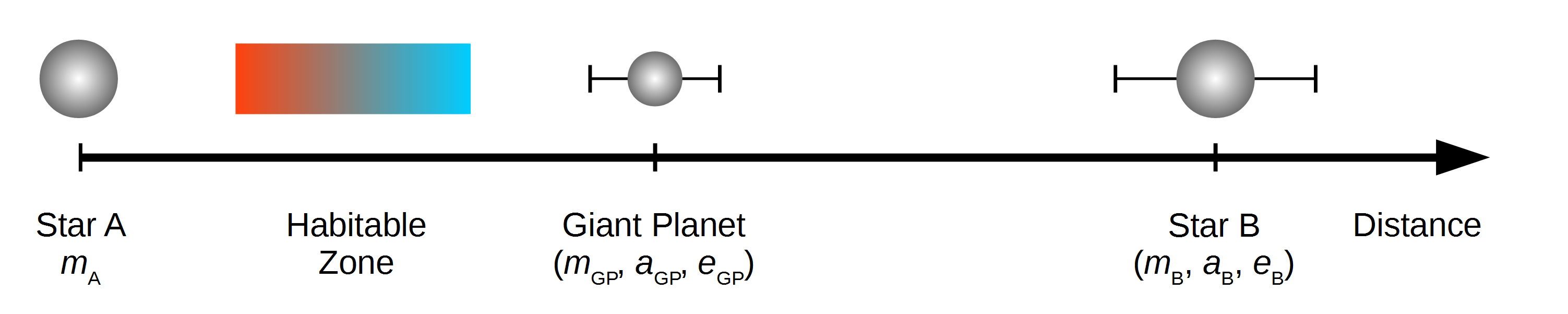}
    \caption{Sketch of the dynamical configuration of a \edit1{coplanar} binary star system. A giant planet (GP) orbits the primary star (star A, or host star) exterior to its habitable zone in circumstellar motion. The more distant secondary star (star B) acts as a perturber to the whole system. Horizontal error bars indicate the radial excursions due to the object's eccentricity. Object sizes and distances are not to scale.}
    \label{fig:dynmodel}
\end{figure}

Let us assume a binary star system that consists of two stellar components and a (Jupiter-like) giant planet (GP), \edit1{with the restrictions that all bodies are in coplanar and prograde motion}.
We will denote the planet hosting star (component A) the primary star, while the distant perturber (component B) will be named the secondary star.
The planet is in circumstellar motion (S-type; \citealt{Dvo1982}) about the host star, and we further assume that it is located exterior to the habitable zone.
Figure \ref{fig:dynmodel} shows a sketch of this situation.
In the figure, the horizontal errorbars indicate the range of distances from pericenter to apocenter that each object can cover due to its eccentricity.
The secondary star must maintain a minimum distance from the primary, in order for the giant planet to remain inside the host star's stability region for all times (see \citealt{Rabl1988}, \citealt{Hol1999}, and \citealt{Pil2002} for details).

A binary star with a circumstellar giant planet forms a hierarchical system.
In terms of dynamical perturbations, the distant secondary star perturbs the giant planet, which itself would pass the perturbations on to any bodies in the HZ.
By virtue of the vastly different masses of the bodies entangled in this interaction, an initially small perturbation from the secondary star can result in huge perturbations in the HZ.
The focus is on secular interactions, where orbital precession frequencies play a major role over extended periods of time, as opposed to short period effects by orbital resonances.

Foremost, the strongest secular perturbations arise \edit1{at locations where the linear combinations of a test planet's orbital precession frequency, $g_{TP}$, and that of the giant planet, $g_{GP}$, form small numbers}
\begin{equation} \label{eq:freqequal}
    q \, g_{TP} - p \, g_{GP} \approx 0,
\end{equation}
\edit1{i.e. when the two frequencies are close to form an integer combination $g_{TP} / g_{GP} \approx p / q$ (with $p, q \in \mathbb{Z}$)}.
This relation forms the basis for all subsequent investigations.
Similar to MMR, the strongest effects arise when $p = q = 1$, \edit1{which gives rise to a linear SR}.
\edit1{The general case with $p \neq q$ ($p > 0, q > 0$)} is called a non-linear secular resonance \citep[see][for the definitions of linear and non-linear SR]{Kne1991}.
\edit1{%
Any given SR's strength depends on the specific frequency combination $(p, q)$.
In case of apsidal frequencies it depends on different powers of the eccentricities,
according to \citet[Appendix~B]{Mur1999} the leading term is $\propto e_{TP}^{q} e_{GP}^{p}$.
So we can appreciate that for fixed $(e_{TP}, e_{GP})$ the SR with $p = q = 1$ will always be dominating, but the $p = 2, q = 1$ and $p = 1, q = 2$ resonances might show very different effects when $e_{TP} \neq e_{GP}$.
}

\edit1{In case of an exact resonance we can recast \eq{eq:freqequal} into the form $q \, g_{TP} = p \, g_{GP}$.}
The right-hand side \edit1{of this equation} constitutes a three-body problem including the two stars and a giant planet; if the latter is taken to be massless, the dynamical model reduces to a restricted three-body problem.
Conversely, the left-hand side inevitably forms a restricted four-body problem, which follows from our convention to set test planet masses $m_{TP} = 0$.
This duality makes it difficult to treat \eq{eq:freqequal} using a single formula or secular model.
An unified formula would need to be applicable to both the three- and four-body problem and to be valid for a large range of parameters with high accurracy.

In fact, up to now our method of choice was the semi-analytical model (SAM) \edit1{described} in \citet{Pil2016} and \citet{Baz2017}.
The SAM starts from the basic equation above and determines the exact location for which $q \, g_{TP} - p \, g_{GP} = 0$ is fulfilled in terms of $a_{TP}$.
It treats the global restricted four-body problem binary-star -- giant planet -- test planet as two coupled three-body problems.
In a first step, the giant planet's precession frequency $g_{GP}$ in \eq{eq:freqequal} is determined numerically by frequency analysis from a numerical simulation of the binary-star -- giant planet subsystem.
Then the calculation of $g_{TP}$ is performed for fixed parameters of the secondary star and giant planet with the determined (and constant) value of $g_{GP}$ by means of any suitable analytical formula.

The trade-off with an approach like the SAM is that it is less flexible when it comes to scan through a multi-dimensional parameter space.
However, a combination of analytical models for both frequencies in \eq{eq:freqequal} would facilitate such studies and allow to investigate a large variety of binary star configurations.
In \citet{Baz2017a} we have investigated a number of analytical secular models for the giant planet precession frequency, $g_{GP}$.
It turned out that two models are promising candidates for this task, namely the models of \citet{And2017} and \citet{Geo2003, Geo2009}.
The latter one provides a framework for the eccentricity evolution (on short and secular time-scales) in the hierarchical full three-body problem, while the former one is only valid for the restricted problem.
In the following, we describe the transition from the SAM to the new Combined Analytical Model (CAM) that involves a combination of two different analytical models that are connected in order to solve \eq{eq:freqequal}.


\subsubsection{Analytical model for giant planet frequency} \label{sec:method:andegg}

\citet{And2017} established a simple and accurate secular model for the restricted three-body problem, i.e. the giant planet acts like a massless body relative to the two stars.
Their model is also applicable in case of strongly perturbed three-body systems, i.e. when the secondary star's distance from the primary is just two to three times that of the giant planet.
The main relation is given by
\begin{equation} \label{eq:andeggmain}
    g_{GP} = g_{H} (1 - \delta_g),
\end{equation}
where $g_H$ is a first order approximation (in masses) to $g_{GP}$ due to Heppenheimer's formula \citep{Hep1978},
\begin{equation} \label{eq:heppenh}
    g_{H} = \frac{3}{4} n_{GP} \left( \frac{m_B}{m_A} \right) \left( \frac{a_{GP}}{a_B} \right)^{3} (1 - e_{B}^{2})^{-3/2},
\end{equation}
and $n_{GP}$ is the giant planet's mean motion.
The term $\delta_g$ is an empirical correction to the frequency, and has the general form
\begin{equation} \label{eq:andeggcorr}
    \delta_g = \sum_{k = 1}^{M} A_k \left( \frac{m_B}{m_A} \right)^{p_k} \left( \frac{a_{GP}}{a_B} \right)^{q_k} e_B^{r_k} ,
\end{equation}
with typically $M \le 20$ coefficients $A_k$ and various powers $(p_k, q_k, r_k)$; refer to \citet{And2017} for the actual values of these coefficients and exponents.
This kind of correction to the secular precession frequency follows the spirit of previous attempts, e.g. \citet{Giu2011}.

The model expresses the secular precession frequency $g_{GP}$ as a polynomial function of the basic parameters mass ratio $\gamma = m_{B} / m_{A}$, semi-major axis ratio $\alpha = a_{GP} / a_{B}$, and perturber eccentricity $e_B$.
Formally, \citet{And2017} state the validity ranges of their fit as $0.1 \le \gamma \le 10$, $0.01 \le \alpha \le 0.4$, and $0 \le e_{B} \le 0.6$.
This interval of mass ratios covers the full range of interest for binary star systems with a G-type primary component, from the lowest mass M-dwarves (for $\gamma = 0.1$) up to F- and A-type stars (for $\gamma \approx 2$).
Larger values of $\gamma$ would lead to an exceedingly fast stellar evolution of the secondary, such that it were no more a main-sequence star.
The interval for the distance ratio $\alpha$ also covers most -- if not all -- currently observed exoplanets in binary systems, including cases like $\gamma$~Cephei ($\alpha = 0.1$) and HD~196885 ($\alpha = 0.12$), see \citet{Baz2017} for a list of tight binary systems.
One remaining caveat is that for some observed binary star systems the secondary's eccentricity might be larger than $e_{B} = 0.6$ (e.g. for $\tau$~Bootis)\footnote{Another issue stems from the fact that the Washington Double Star catalogue lists less than 1000 systems (from a total of more than 130,000) with well-constrained stellar eccentricity. At larger stellar separations (above 100 au, i.e. in the distance range considered here) there is a deficit of observational information on the eccentricities. Therefore, there is a clear need to estimate perturbing effects on the HZ due to stellar eccentricities.}.


\subsubsection{Analytical model for test planet frequency} \label{sec:method:laplag}

According to the Laplace-Lagrange model (LLM) the secular precession frequency for a massless test planet (TP) influenced by $N$ massive perturbers is given by
\begin{equation*}
    g_{TP} = \frac{n_{TP}}{4} \sum_{i=1}^{N}
             \left( \frac{m_{i}}{m_{A}} \right)
             \left( \frac{a_{TP}}{a_{i}} \right)^{2}
             b_{3/2}^{(1)} \left( \frac{a_{TP}}{a_{i}} \right)
            (1 - e_{i}^{2})^{-3/2},
\end{equation*}
where $n_{TP}$ is the test planet's mean motion, and the function $b_{3/2}^{(1)} (\alpha_i)$ is a Laplace coefficient of the single scalar variable $0 \le \alpha_{i} = a_{TP} / a_{i} < 1$ \citep[cf.][chapter~6]{Mur1999}.
For our dynamical model with $N = 2$ (giant planet and secondary star) it follows that
\begin{equation} \label{eq:laplag}
    \begin{split}
        g_{TP} = \frac{n_{TP}}{4} &
        \left[
            \left( \frac{m_{GP}}{m_{A}} \right)
            \left( \frac{a_{TP}}{a_{GP}} \right)^{2}
            b_{3/2}^{(1)} \left( \frac{a_{TP}}{a_{GP}} \right)
            (1 - e_{GP}^{2} )^{-3/2}
        + \right. \\
        & \left. +
            \left( \frac{m_{B}}{m_{A}} \right)
            \left( \frac{a_{TP}}{a_{B}} \right)^{2}
            b_{3/2}^{(1)} \left( \frac{a_{TP}}{a_{B}} \right)
            (1 - e_{B}^{2})^{-3/2}
        \right].
    \end{split}
\end{equation}
Note that this is an ad-hoc modification of the classical LLM.
Originally, the classical secular theory is of first-order in the eccentricity and hence is valid only for low values of $e_{i}$.
However, by adding the extra terms $(1 - e_{i}^{2})^{-3/2}$ with the eccentricity dependence we can extend its range of validity.
The motivation is that \eq{eq:laplag} looks and behaves similar to \eq{eq:heppenh} with the additional terms added.
In fact, \citet{Baz2017a} have shown that such a modification acts as an interpolating formula between the classical LLM (in the limit $e_{i} \rightarrow 0$) and the Heppenheimer model (for $a_{TP} / a_{i} \rightarrow 0$).

Equation (\ref{eq:laplag}) can handle arbitrary mass ratios and is applicable for large distance ratios up to the limit $a_{TP} / a_{i} \rightarrow 1$, which becomes important when the giant planet approaches the outer HZ border (OHZ), but we strictly require that $\ahzout < a_{GP}$.
Alternatively, the LLM can be replaced by any suitable secular model that can handle at least a four-body problem, e.g. the model of \citet{Mil1990}.


\subsubsection{Combined Analytical Model (CAM)} \label{sec:method:cam}

After having introduced both analytical sub-models, we now combine equations (\ref{eq:andeggmain}) and (\ref{eq:laplag}) to express \eq{eq:freqequal} in terms of the five parameters $(m_{GP}, m_{B}, a_{GP}, a_{B}, e_{B})$.
We assume that the eccentricity of the giant planet ($e_{GP}$) is small enough to neglect its effect, i.e. $e_{GP} \rightarrow 0$, which reduces the dynamical system by this variable.

In order to have a more compact notation, we introduce the following abbreviations: $\gamma_{1} := m_{GP} / m_{A}$, $\gamma_{2} := m_{B} / m_{A}$, $\alpha_{1} := a_{TP} / a_{GP}$, $\alpha_{2} := a_{GP} / a_{B}$, so that we can express $a_{TP} / a_{B} = (a_{TP} / a_{GP}) (a_{GP} / a_{B}) = \alpha_1 \alpha_2$.
Furthermore, the eccentricity $e_{B}$ appears on both sides of the equation in form of the expression $\epsilon_{B} := (1 - e_{B}^{2})^{-3/2}$.

With all these definitions we can write \eq{eq:freqequal} \edit1{for linear\footnote{Without loss of generality we focus here on linear secular resonances, because these show the maximum effect in terms of forced eccentricities (cf. Fig.~\ref{fig:emaxhz}). In a more general setting, the coefficients $(p, q)$ can be easily reintroduced into \eq{eq:freqeqmain} to study also nonlinear SR (see Fig.~\ref{fig:nonlinsr}).} SR ($p = q = 1$)} in terms of $(\gamma_{1}, \gamma_{2}, \alpha_{1}, \alpha_{2}, \epsilon_{B})$ to obtain
\begin{equation} \label{eq:freqeqmain}
    \begin{split}
        \frac{1}{4} n_{TP}
        \left(
            \gamma_{1} \alpha_{1}^{2} b_{3/2}^{(1)} (\alpha_{1}) +
            \gamma_{2} \left( \alpha_{1} \alpha_{2} \right)^{2} b_{3/2}^{(1)} (\alpha_{1} \alpha_{2}) \epsilon_{B}
        \right) =\\
        \frac{3}{4} n_{GP} \gamma_{2} \alpha_{2}^{3} \epsilon_{B} (1 - \delta_g).
    \end{split}
\end{equation}
We will refer to \eq{eq:freqeqmain} from now on as Combined Analytical Model (CAM).

The expression for $\delta_g$ is given in \eq{eq:andeggcorr} as a function of $(\gamma_{2}, \alpha_{2}, e_{B})$, where $e_{B}$ appears explicitly.
From now on we skip the subscript $g$ from $\delta_g$, and are going to use simply $\delta$ when referring to \eq{eq:andeggcorr}.
Note that the frequencies (mean motions) $n_{TP} = n(a_{TP})$ and $n_{GP} = n(m_{GP}, a_{GP})$ are the only dimensional quantities in this equation, and that they are known functions of the system parameters.


\subsubsection{Derivation of simplified CAM} \label{sec:method:simpcam}

Naturally, it would be helpful to have at hand some sort of analytical expressions that allow to investigate the qualitative behaviour of the CAM solutions.
These expressions are not available in general, but they can be found for an approximate form of \eq{eq:freqeqmain} that we will derive now.

The primary obstacle in \eq{eq:freqeqmain} are the Laplace coefficients $b_{3/2}^{(1)}$.
We can get rid of them by assuming that the conditions $\alpha_{1} \ll 1$ and $\alpha_{2} \ll 1$ hold, such that we can replace these functions by their MacLaurin series expansions
\[
    b_{3/2}^{(1)}(\alpha) \approx 3 \alpha + \frac{45}{8} \alpha^{3} + \mathcal{O}(\alpha^{5})
\]
up to the lowest order in $\alpha$.
This approximation is justified because of the hierarchical nature of the dynamical system.
In case of well separated orbits we expect quite small values of $\alpha < 0.1$, for which this approximation yields relative errors of roughly $1 - 2$~\% compared to the values evaluated numerically from the full form of the Laplace coefficient.
Deviations of this magnitude are tolerable given the inherent fit-errors of the secular model in \eq{eq:andeggmain}, that itself has median errors of $2 - 3$~\% relative to a fully numerical solution (compare section \ref{sec:result:modelacc}).

Starting from \eq{eq:freqeqmain}, we replace the Laplace coefficients by their approximations, insert $n_{TP} = (G m_{A} / a_{TP}^{3})^{1/2}$ and $n_{GP} = (G m_{A} (1 + \gamma_{1}) / a_{GP}^{3})^{1/2}$, and eliminate common factors on both sides of the equation to obtain
\begin{equation} \label{eq:freqeqapprox}
    \gamma_{1} \alpha_{1}^{3} + \gamma_{2} \alpha_{1}^{3} \alpha_{2}^{3} \epsilon_{B} = (1 + \gamma_{1})^{1/2} \gamma_{2} \alpha_{1}^{3/2} \alpha_{2}^{3} \epsilon_{B} (1 - \delta).
\end{equation}
This equation, termed the simplified CAM, represents an approximate form of \eq{eq:freqeqmain} that shares all of its properties in the limit $\alpha_{1,2} \rightarrow 0$.
Appendix~\ref{sec:app:scam} provides (approximate) analytical solutions to this simplified CAM with respect to the parameters $(\gamma_{1}, \gamma_{2}, \alpha_{1}, \alpha_{2}, \epsilon_{B})$.


\subsection{Properties of the CAM} \label{sec:method:propcam}

We are now able to investigate the binary-star -- giant planet -- test planet system as a coupled dynamical problem given by \eq{eq:freqeqmain} depending on five parameters.
The goal is to identify combinations of those parameters that would lead to secular perturbations inside the HZ, i.e. for $\ahzin \le a_{TP} \le \ahzout$.
This knowledge is useful to exclude such systems from searches for potentially habitable planets, and allows for a quick dynamical classification of the HZ (resonant vs non-resonant).


\subsubsection{Searching for CAM solutions} \label{sec:method:solution}

Our approach is to find two sets of solutions of \eq{eq:freqeqmain}: in the first case for $a_{TP} = \ahzin$, and in the second case by setting $a_{TP} = \ahzout$.
The two solutions obtained in this way mark the extremal cases where the perturbation just touches the HZ borders.
For parameter values intermediate to the two identified sets the perturbation will fall somewhere into the HZ.
However, one must be aware that there are circumstances for which these solutions are still too conservative.
An example is that the SR falls just slightly outside of the HZ, such that our model would predict that it does not have any influence.
Resonances have a certain resonance width in real systems though, this width depends on the strength of the perturbation \citep[see, e.g.][]{Mal1998}.
So, if the SR were located slightly outside the HZ and had a resonance width that is sufficiently large then it would reach into the HZ and does affect the dynamics there.
Such situations must be carefully investigated on an individual basis; often numerical simulations will be necessary to decide whether or not a resonance near the HZ has an influence on it \citep[see also][Fig.~2]{Baz2017}.

To simplify the search for solutions to the CAM in the five-dimensional parameter space, we fix 4 of the 5 variables and consider \eq{eq:freqeqmain} to be a one-dimensional function of the remaining variable.
We then solve the equation via an iterative process, because it depends in a non-trivial way on the variables $\alpha_{1}$ and $\alpha_{2}$ (via the Laplace coefficients).
This procedure yields parametric solutions for the selected variable as a function of the other four parameters, and the solutions can be visualized in a two-dimensional parametric space like shown in Fig.~\ref{fig:motiv} and others in section \ref{sec:result}.


\subsubsection{A motivating example} \label{sec:method:motivation}

\begin{figure}
    \centering
    \includegraphics[width=0.5\textwidth]{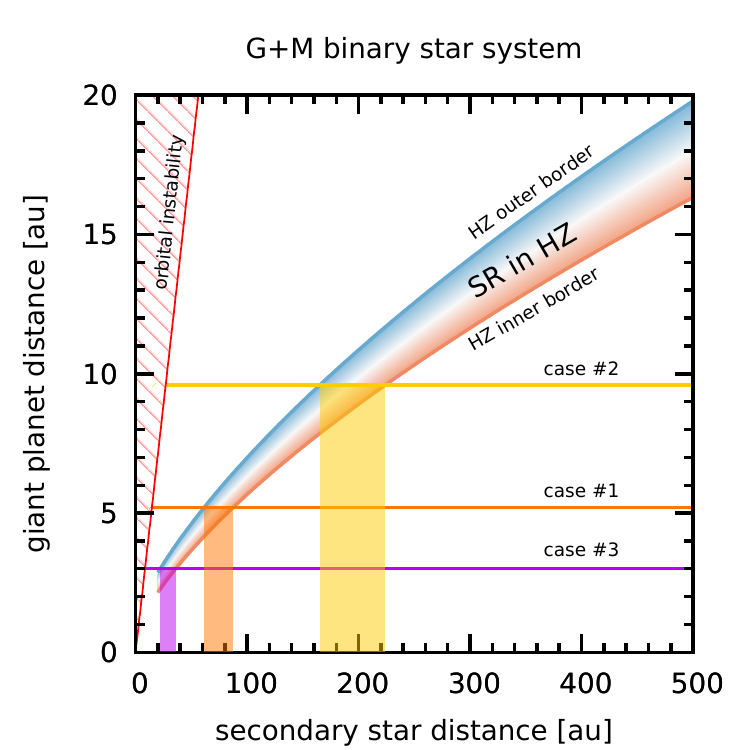}
    \caption{Along the curved diagonal strip an SR affects some part of the HZ, while in the surrounding white area there is no SR inside the HZ. The top (blue) curve marks the outer HZ border, and the lower (red) curve delimits the inner HZ border. Three example cases show parameter combinations for a fixed giant planet distance (horizontal lines): 5.2~au for case~1 (orange), 9.6~au for case~2 (yellow), \edit1{or 3.0~au for case~3 (purple)}. For each example a certain range of secondary star distances (coloured boxes) induces an SR in the HZ. The hatched area on the left margin indicates the critical distance \edit1{beyond which orbital instability} of the planet is expected.}
    \label{fig:motiv}
\end{figure}

We present \edit1{as motivation a first} application of the CAM in Fig.~\ref{fig:motiv}.
In this example a binary star system is depicted that consists of a G-type primary and an M-type secondary star.
The main feature is a diagonal stripe (bottom left to top right) that represents the set of parameters in the ($a_{B}, a_{GP}$) parameter space that lead to an SR in the HZ.
\edit1{Three} specific examples serve to highlight the strength of this approach.
In the first case (medium grey or orange colour) a Jupiter mass planet at 5.2~au (\edit1{Jupiter's actual semi-major axis in the solar system}) would perturb the HZ if the secondary was at a distance between $60-80$~au (see filled rectangle in the middle).
The second case (light grey or yellow colour) places a Jupiter mass planet at Saturn's position at 9.6~au, where this planet induces an SR in the HZ for stellar distances of roughly $170-220$~au (lighter filled rectangle to the right).
\edit1{%
Another case shows the same Jupiter-mass planet at a distance of only 3~au (dark grey or purple colour) with perturbations occurring in the HZ for stellar distances between $22 - 36$~au (darker filled rectangle to the left).
}
Note that the band is \edit1{artificially truncated on both ends;} for $a_{B} < 20$~au, because such a tight binary system would require to place the GP below 2~au \edit1{to maintain its orbital stability for higher secondary eccentricities, but in turn} this limits orbital stability in parts of the HZ.
In practice, the stripe could be traced out to much larger secondary star (and planet) distances, but the more distant the giant planet becomes the weaker its influence on the HZ would be.


\subsubsection{Effect of a secular perturbation in the HZ} \label{sec:method:emax}

\begin{figure*}
    \centering
    \includegraphics[width=0.3\textwidth,page=1]{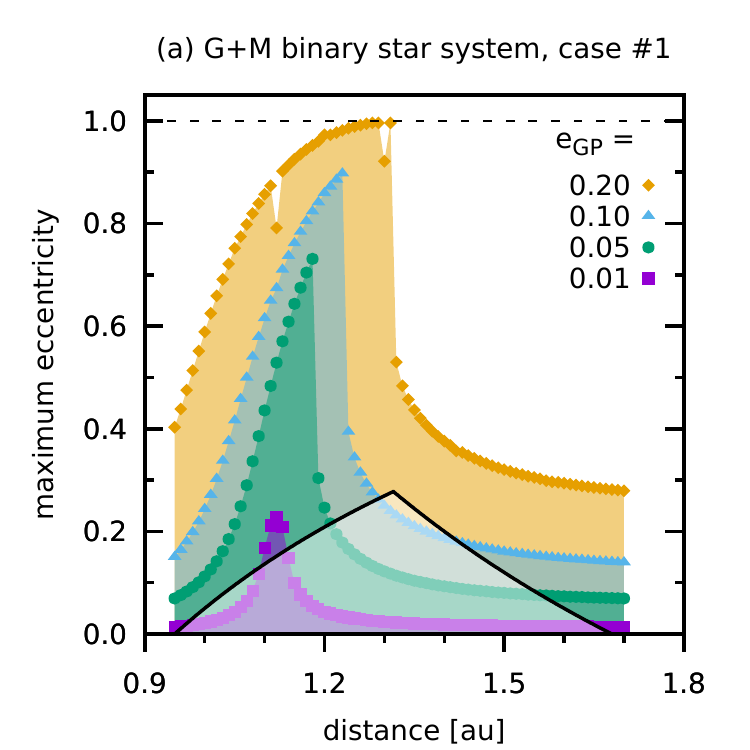}
    \includegraphics[width=0.3\textwidth,page=2]{fig3.pdf}
    \includegraphics[width=0.3\textwidth,page=3]{fig3.pdf}
    \caption{Maximum eccentricities of test planets distributed across the HZ. The $y$-axis gives the maximum value of the eccentricity (max-e) over the simulation time span of 2~Myrs. Different symbols (and colours) distinguish specific initial eccentricities of the perturbing giant planet. \edit1{The broad peak in max-e in each panel corresponds to the location of the linear SR with the GP; narrower peaks represent locations of MMR.} Panels (a)--(c) refer to the respective example cases 1--3 from Fig.~\ref{fig:motiv}. A triangle-like overlay indicates the permanent HZ as a function of the test planet's eccentricity.}
    \label{fig:emaxhz}
\end{figure*}

Apart from being able to locate and identify secular perturbations some important questions remain:
How much does it matter? Would potential planets in the HZ even care about such perturbations?
The answer is likely yes!
To understand the reason, we need to consider the driver of habitable conditions: the stellar irradiation.

It became evident from the investigations of \citet{Eggl2012}, \citet{Kal2013}, \citet{Cun2014}, and others that a static definition of habitable zones in binary star systems is only a rough approximation, and for some applications it is unjustified.
The gravitational interaction of the (at least) three bodies results in quasi-periodic oscillations of the planet's eccentricity even for initially circular planetary orbits.
These oscillations in eccentricity are correlated with phases of larger amplitude variations in irradiation.
Consequently, a planet that may sustain habitable conditions on a low eccentricity orbit could leave the permanently habitable zone (PHZ) for a temporarily increased eccentricity -- just to return to a moderate eccentricity some time later.
It is this effect on the eccentricity that effectively limits the habitability.

In Fig.~\ref{fig:emaxhz} we provide the results of calculations for (massless) test planets distributed across the HZ for the \edit1{three} cases from Fig.~\ref{fig:motiv}.
\edit1{These fully numerical simulations were performed with the N-body code Mercury \citep{Cha1999} using the adaptive time-step Radau method with local error tolerance $10^{-12}$.}
Each figure shows the maximum eccentricities (max-e) of test planets as a function of their initial location.
The max-e is color coded according to four different scenarios related to the giant planet's orbital eccentricity (from bottom to top): (1) $e_{GP} = 0.01$ (square symbols), (2) $e_{GP} = 0.05$ (circles), (3) $e_{GP} = 0.1$ (triangles), and (4) $e_{GP} = 0.2$ (diamond symbols).
Panel (a) for case~1 refers to the giant planet at 5.2~au, in panel (b) the planet is located at 9.6~au, while in panel (c) the planet is at 3.0~au.
A triangle-shape overlay (in light gray) delineates the region of the PHZ for the test planets.
Any test planet that acquires a max-e value outside of this region must be considered uninhabitable (at least temporarily).

In the first example case the perturbing giant planet causes a linear SR near $1.1 - 1.2$~au that is marked by the spikes in max-e.
Even for $e_{GP} = 0.01$ the GP forces test planets locally to max-e values of 0.2.
An increase of $e_{GP}$ to 0.05 then boosts the eccentricities to values beyond 0.6.
In these latter cases some small intervals with moderate max-e remain, e.g. from $1.2 - 1.55$~au ($1.3 - 1.4$~au) for $e_{GP} = 0.05$ (0.1), respectively.
For even higher $e_{GP}$ no test planet can maintain max-e values low enough to stay in the PHZ; some even escape from the system on unbound trajectories.

The second example with a more distant GP in panel (b) exhibits a linear SR at about the same location as before.
The spike in max-e is not as pronounced as in the other example, but it is still visible even for relatively small $e_{GP}$.
An increase in $e_{GP}$ first leads to the deterioration of habitability near the edges of the HZ, but increasingly also in the inner regions.
For $e_{GP} = 0.1$ already many test planets achieve too high max-e values and only in an interval from $1.25 - 1.5$~au they would remain permanently habitable; for an even more eccentric GP no habitable test planet remain any more.

\edit1{%
Case~3 in panel (c) represents a strongly perturbed HZ by a GP at a distance of only 3~au.
Besides the main linear SR at about 1.1~au also numerous narrow peaks appear that correspond to MMR with the GP, see Table~\ref{tab:mmr} for a list of MMR.
The width and max-e amplitude of the SR are strongly modified when increasing the GP's eccentricity, such that for $e_{GP} > 0.1$ no habitable test planets remain.
Note that apparently the 5:1 MMR has the largest effect (max-e $> 0.9$ already for $e_{GP} \ge 0.05$), while the theoretically stronger 3:1 and 5:2 MMR are visible only for $e_{GP} = 0.1$ or higher.
Although the 5:1 MMR is a fourth order MMR, here it interacts with the SR and becomes the dominant MMR locally, apart from the 2:1 MMR that is located slightly beyond the right edge of this figure.
}

\begin{table}
    \centering
    \caption{A selection of relevant MMR in the HZ for $a_{GP} = 3$~au (case~3).}
    \label{tab:mmr}

    \begin{tabular}{r@{:}lc}
      \toprule
    \multicolumn{2}{c}{MMR} & Location [au] \\
      \midrule
    5&1  & 1.03 \\
    9&2  & 1.10 \\
    17&4 & 1.14 \\
    4&1  & 1.19 \\
    3&1  & 1.44 \\
    8&3  & 1.56 \\
    5&2  & 1.63 \\
    2&1  & 1.89 \\
      \bottomrule
    \end{tabular}
\end{table}

These examples show clearly that the GP's eccentricity plays a crucial role for secular perturbations in the HZ.
\edit1{It is obvious that slight variations in $e_{GP}$ can result in large perturbations that involve the large forced max-e of the TP induced by low to moderately eccentric GP, and the feedback effects on $e_{GP}$ discussed in section \ref{sec:result:planpar}.}

\begin{figure}
    \centering
    \includegraphics[width=0.5\textwidth]{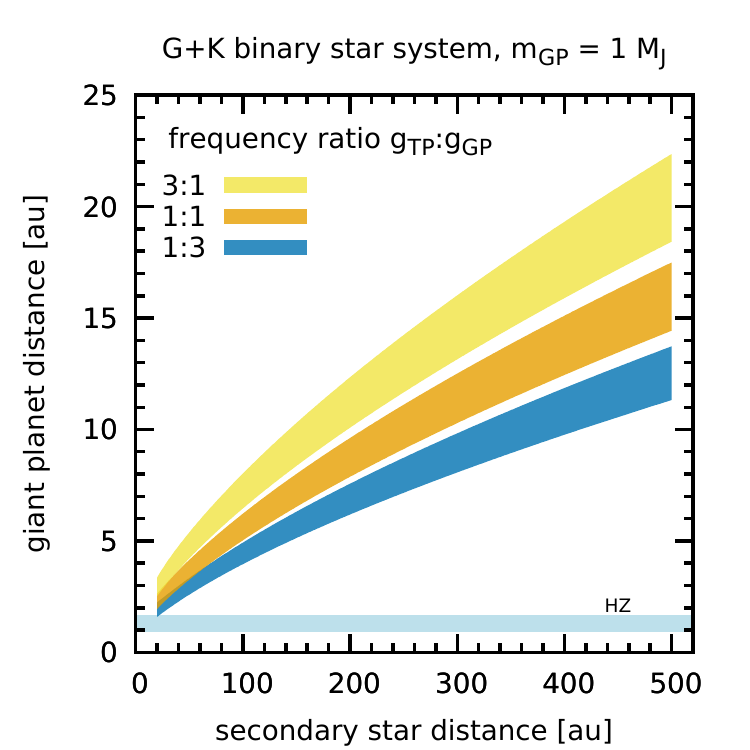}
    \caption{\edit1{Examples of nonlinear SR locations identified via CAM. From top to bottom the shaded bands represent the following SR: the $g_{TP} - 3 \, g_{GP}$, $g_{TP} - g_{GP}$ (linear), and $3 \, g_{TP} - g_{GP}$. The horizontal stripe at the bottom represents the habitable zone.}}
    \label{fig:nonlinsr}
\end{figure}

\edit1{%
Figure~\ref{fig:nonlinsr} depicts a binary star system with a K-type secondary ($m_{B} = 0.7 \, M_{\odot}$) and the bands of various SR.
Besides the linear SR (middle curve) the figure also shows the $(q, p) = (1, 3)$ (top curve) and $(q, p) = (3, 1)$ (bottom curve) nonlinear SR.
Note that the vertical width of the curves in $a_{GP}$ does not indicate their strength (nonlinear SR are rather weak, cf. Fig.~\ref{fig:laplagerr}), but rather the feasible range of GP distances that induce these SR in the HZ.
At small secondary star distances these three curves overlap which means that two (or more) SR may appear simultaneously in the HZ.
}




\section{Results} \label{sec:result}

\edit1{In a first part (section \ref{sec:result:modelacc}) we assess the analytical model's accuracy (i.e. the predicted secular frequencies $g_{GP}$ and $g_{TP}$) relative to numerical reference results, while in section \ref{sec:result:apply} we apply the CAM to different binary systems and present the effect of varying different system parameters.}


\subsection{Model accuracy} \label{sec:result:modelacc}

\edit1{%
As the full CAM consists of two models we check the performance of each of them independently.
Section \ref{sec:result:camgp} contains the checks related to the CAM and $g_{GP}$, while section \ref{sec:result:camtp} contains results pertaining to the $g_{TP}$ part.
Additionally, we compare the results of the full CAM (\eq{eq:freqeqmain}) to those of the simplified CAM (\eq{eq:freqeqapprox}) in section \ref{sec:result:simpcam}.
}


\subsubsection{Accuracy of full CAM: GP frequency} \label{sec:result:camgp}

We performed a series of test computations on a two-dimensional grid in the $(a_{B}, e_{B})$ parameter space for four different masses of the secondary star.
The main goal was to assess the accuracy of the analytical model in \eq{eq:andeggmain} when its assumptions of zero mass and eccentricity for the GP are not met.
In fact, we tested several different GP masses and non-zero eccentricities for a total of more than 14,000 test runs.
\edit1{%
The simulations were done in the framework of the full three-body problem, purely Newtonian gravity, and the equations of motion were solved with the Radau method from the Mercury package \citep{Cha1999}.
The secular peridos vary over several orders of magnitude (roughly $10^3 - 10^7$ years) in the parameter range of interest, so we made sure to cover a simulation time of at least 20 secular periods (estimated from \eq{eq:heppenh}) and to obtain at least 2000 data points for the time evolution of the heliocentric orbital elements.
}
Each of these individual test runs was subject to a frequency analysis to obtain a reference value for $g_{GP}$.
\edit1{The reference frequency was obtained by the FMFT method \citep{Sid1996, Las1999} from the time series of the complex variable $z(t) = h(t) + i k(t)$, where $i = \sqrt{-1}$, with $h(t) = e(t) \sin \varpi(t)$ and $k(t) = e(t) \cos \varpi(t)$.}

\begin{figure}
    \centering
    \includegraphics[width=0.5\textwidth]{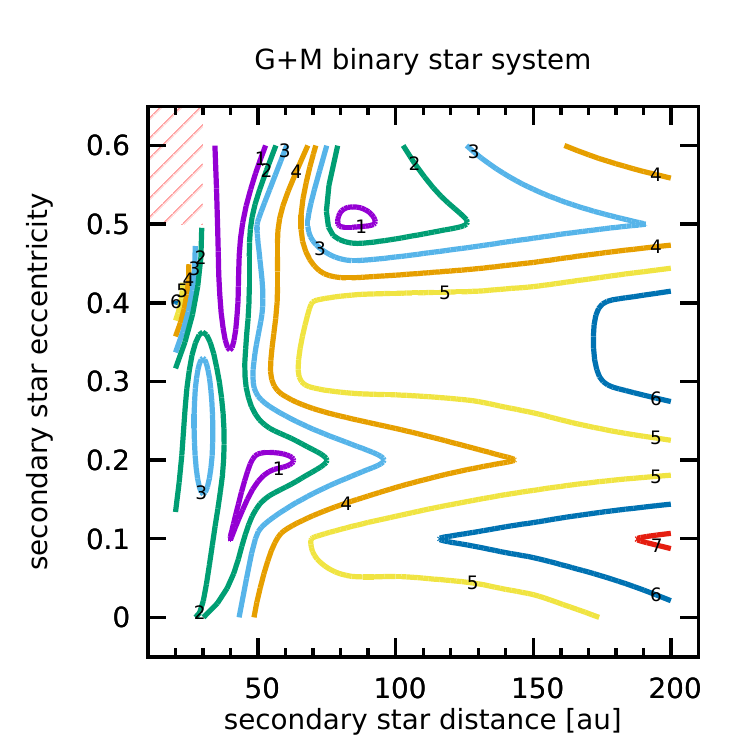}
    \caption{Comparison of the deviations between the analytical model in \eq{eq:andeggmain} and results from fully numerical simulations for a G+M binary star system. The contour lines trace the relative errors (in percent) of the giant planet's precession frequency $g_{GP}$. In the top left corner (hatched area) the giant planet's orbital instability precludes a comparison.}
    \label{fig:contours}
\end{figure}

Figure~\ref{fig:contours} displays a comparison of the CAM and the numerical reference values.
The contours show the relative error in percent for the case of an M-type secondary, where the giant planet is located at 3~au on an initially circular orbit.
In this case the agreement is very good, with a maximum deviation of less than 8~\%.
Generally, we observe the largest errors in the top-left corner for tight and highly eccentric secondaries, where orbital instability of the GP becomes an issue (cf. hatched area).
When increasing the giant planet's mass to 5~$M_J$ also the maximum error increases to $\sim 30$~\%, while the average error stays below 10~\%.

\begin{figure}
    \centering
    \includegraphics[width=0.5\textwidth]{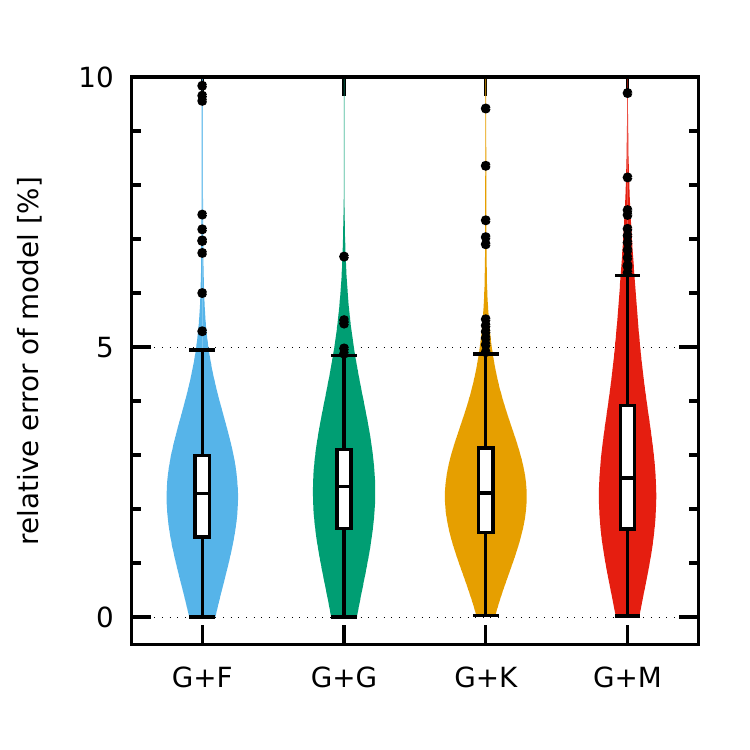}
    \caption{Overview of the relative error (on the $y$-axis) between the analytical model and fully numerical simulations for all four binary cases. The boxes represent the range in between the first and third quartiles, with a horizontal line at the median value. The errorbars indicate the range that spans 95~\% of all points in the data set; outlying points beyond these limits are visible as black dots at larger errors. Shaded regions show the normalized frequency distribution of data points obtained via a kernel density estimate using Gaussian kernels.}
    \label{fig:andstats}
\end{figure}

We show a more global comparison in Fig.~\ref{fig:andstats}, that gives an overview on the statistical distributions of the relative errors for the four secondary star masses.
For this figure we selected the subset of the simulations with $m_{GP} = 1 \, M_{J}$ and distances $a_{GP}$ of 3, 5, and 7~au.
The vertical axis shows that the relative errors are usually quite small with only few exceptions.
For each of the four cases a boxplot indicates the range of errors in between the first and third quartiles, horizontal lines represent the median values that lie around 2.5~\% in all cases.
Furthermore, the region enclosed by the errorbars contains 95~\% of the data points, while black dots mark individual outliers.
The normalized frequency distributions of data points are shown as shaded regions.
It is thus clearly visible that the distributions are all unimodal and that in $> 95$~\% of cases the analytical model deviates by less than 10~\% from the numerical reference value.
The remaining outliers stem mainly from those simulations that combine a highly eccentric and tight secondary with a more distant giant planet.
These edge cases lie close to the instability border (see top left corner in Fig.~\ref{fig:contours}) and are characterized by large values of $\alpha_{2}$.


\subsubsection{Accuracy of full CAM: TP frequency} \label{sec:result:camtp}

\edit1{%
A second set of test simulations covers \eq{eq:laplag}.
The aim is to assess how well this equation performs relative to numerical reference results in the presence of two massive perturbers.
All simulations are performed in a restricted four-body problem with the Radau method (see section~\ref{sec:method:emax} for details) and cover a duration of 2~Myr, with output every 100 years in order to resolve short period variations in eccentricity.
We obtain the reference frequencies again with the FMFT method.
}

\begin{figure}
    \centering
    \includegraphics[width=0.5\textwidth]{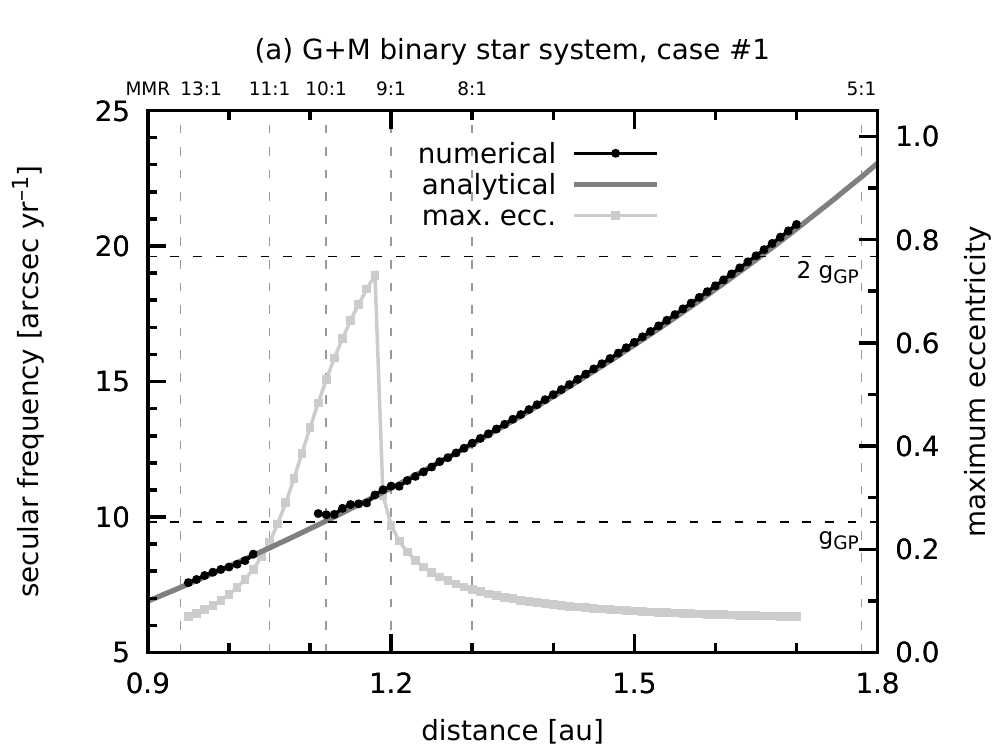}\\
    \includegraphics[width=0.5\textwidth]{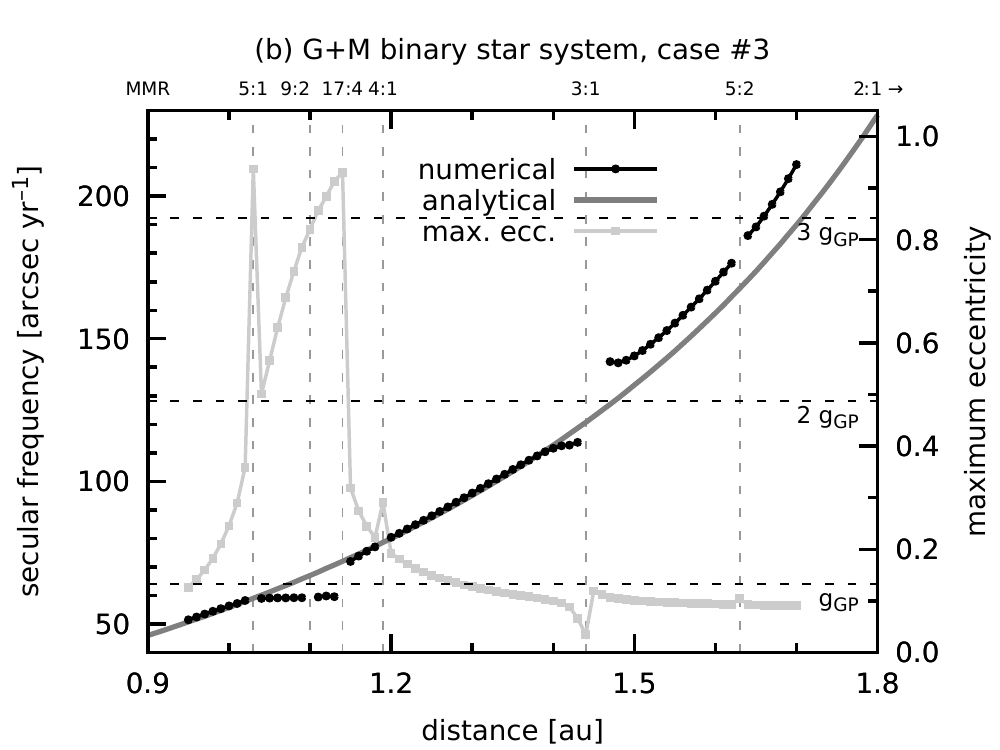}
    \caption{\edit1{Comparison of secular frequencies obtained from the analytical model in \eq{eq:laplag} with frequencies from fully numerical simulations for two G+M binary star systems. Panel (a) corresponds to case~1 from Fig.~\ref{fig:motiv} with $a_{GP} = 5.2$~au, while panel (b) is for case~3 with $a_{GP} = 3$~au; initially $e_{GP} = 0.05$ in both cases. The solid grey lines show the analytical frequencies overplotted with black dots for the numerical frequency values (left $y$-axis). As a visual aid we replot the respective max-e curves (light grey squares, right $y$-axis) from Fig.~\ref{fig:emaxhz}. Vertical dashed lines indicate locations of MMR.}}
    \label{fig:laplagerr}
\end{figure}

\edit1{%
Figure~\ref{fig:laplagerr} presents the comparisons for two sample cases.
Panel (a) shows case~1 from Figs.~\ref{fig:motiv} and \ref{fig:emaxhz}, where we have $a_{GP} = 5.2$~au, $e_{GP} = 0.05$, and $e_{B} = 0.01$.
The solid grey curve gives the secular frequencies as predicted by \eq{eq:laplag} (left $y$-axis).
Black dots represent numerically determined apsidal precession frequencies for test planets in the HZ range ($0.95 - 1.70$~au).
We overplot the max-e curve (with light grey squares, see values on the right $y$-axis) to aid the identification of relevant features.
Horizontal dashed lines indicate the GP's proper precession frequency ($g_{GP}$) and multiples of its value, whereas vertical dashed lines indicate MMR (see respective labels at the top).
It is visible that there is an excellent agreement between the analytical and numerical frequencies, except for the interval between the 11:1 and 10:1 MMR.
The nominal linear SR location is at $1.12$~au where the grey curve equals $g_{GP}$; the nonlinear SR $g_{TP} - 2 \, g_{GP} \approx 0$ lies near to $1.65$~au.
According to the numerical data it seems that the exact SR location is slightly shifted to smaller distance ($\lesssim 1.1$~au).
Note that the peak in max-e does not coincide with the location of the SR, moreover it is asymmetric with a sharp edge short of the 9:1 MMR.
A similar qualitative behaviour has also been demonstrated for secular resonances in the solar system \citep[Fig.~3]{Mal1998}.
The max-e curve is more symmetric for low GP eccentricity (cf. Fig.~\ref{fig:emaxhz}), but the asymmetry becomes more pronounced with increasing $e_{GP}$.
}

\edit1{%
The second panel (b) depicts case~3, with the same parameters as above except for $a_{GP} = 3$~au.
Again the light grey points represent the max-e values for individual test particles, where we now have several narrow peaks due to MMR besides the dominating broad SR.
The analytical and numerical frequencies are in rather good agreement, except at locations of MMR and starting from about 1.5~au outwards to larger distances.
A main cause of this disagreement is the influence of the strong 2:1 MMR at 1.89~au (outside of the plot range).
The analytical model in \eq{eq:laplag} only accounts for secular terms of the disturbing function, but here also the resonant (short period) terms play an increasingly important role.
An estimate of the libration width of the 2:1 MMR gives $\delta a \approx 0.06$~au \citep{Mal1998, Mur1999}, but this is not wide enough to reach into the HZ, as we can also see from the flat part of the max-e curve at distances larger than 1.5~au.
The width of the linear SR (at 1.1~au) roughly corresponds to the interval between the 5:1 and 17:4 MMR, compare Tab.~\ref{tab:mmr} for their exact locations.
In this region the test planet's proper precession frequency does not change.
A similar behaviour occurs near the 3:1 MMR that falls into the vicinity of the $g_{TP} - 2 g_{GP}$ nonlinear SR.
Such flat parts (or plateaus) of the frequency curve correspond to crossing of resonance islands, as demonstrated for different MMR in \citet[Fig.~1]{Rob2001}.
Another interesting case is the 3:1 MMR, where the max-e curve shows a `pulse-like' effect.
\citet{Chr1999} found a similar behaviour when they analysed the effect of near MMR on the secular precession frequency.
An explanation for the pulse is that the secular frequency profile is modified by the sign-change of the MMR's small divisor, which -- in case of second-order MMR -- gives a negative contribution before the resonance and a positive one afterwards.
This effect on the secular frequency directly relates to the max-e curve where it is visible, too.
}

\edit1{%
In general, missing data points in Fig.~\ref{fig:laplagerr} indicate a complex dynamical behaviour of the respective particle, where the FMFT was unable to determine a unique dominating frequency.
Note that in a purely secular setting the d'Alembert rules require additional frequencies (like nodal precession frequencies) to satisfy the resonance relation in \eq{eq:freqequal} for a nonlinear SR.
Alternatively, a nonlinear SR may couple with a nearby MMR to fulfill the exact resonance condition, such as for the $(p, q) = (2, 1)$ SR and the 3:1 MMR.
This coupling gives rise to the max-e peaks visible in Fig.~\ref{fig:emaxhz}~(c) for $e_{GP} \ge 0.1$, but also a local minimum in max-e is associated to the same 3:1 MMR.
The presence of SR and nearby MMR does not necessarily lead to adverse effects, because MMR can also have a stabilizing effect.
One particular example was shown in \citet[Fig.~1]{Pil2005} for the $\gamma$~Cephei system, where a stable island around the 3:1 MMR (near to an SR) persists also for non-coplanar motion up to an inclination of $15^{\circ}$ relative to the GP.
}

\edit1{%
Overall, Figs.~\ref{fig:emaxhz} and \ref{fig:laplagerr} clearly show that the linear SR is the dominant perturbation and that higher order SR play a secondary role.
This finding is backed-up by the results of \citet[Fig.~3]{Ban2016} whose max-e plots also show a dominant linear SR that locally couples to various MMR.
Nonlinear SR become relevant for much higher $e_{GP}$ when the resonance widths of SR and MMR are wide enough to overlap and cause extended regions of chaotic motion.
Such an example was presented in \citet{Baz2017} for the HD~196885 system\footnote{Radial velocity observations of the system suggest an exoplanet with minimum mass $3 \, M_{J}$ and eccentricity $e_{GP} \sim 0.5$.} where the nonlinear SR $2 \, g_{TP} - g_{GP}$ and $3 \, g_{TP} - g_{GP}$ overlap with nearby MMR.
For another example of nonlinear SR see Fig.~\ref{fig:nonlinsr}.
}


\subsubsection{Accuracy of simplified CAM} \label{sec:result:simpcam}

Figures \ref{fig:contours} and \ref{fig:andstats} demonstrate that the full CAM provides an accurate description of the secular dynamics.
The open question is how well the simplified CAM reproduces results of the full equation.

\begin{figure}
    \centering
    \includegraphics[width=0.5\textwidth,page=1]{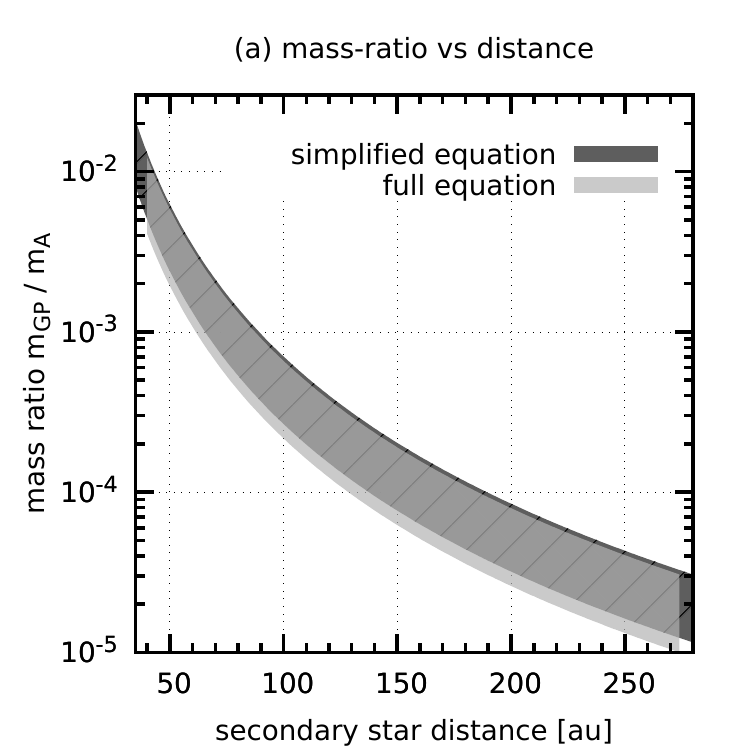}
    \includegraphics[width=0.5\textwidth,page=2]{fig8.pdf}
    \caption{Comparison of model accuracy between simplified CAM (darker colour with pattern in the background) and full CAM (lighter colour). Panel (a) shows the mass ratio $\gamma_{1}$ as a function of secondary star distance for a G+M binary on circular orbits. The plot in panel (b) shows the distance ratio $\alpha_{2}$ as a function of secondary star mass for a Jupiter-mass planet at 5~au. Note that the curves overlap each other at the bottom border, but the curve for full CAM is slightly more extended at the upper border.}
    \label{fig:scam-vs-fullcam}
\end{figure}

A first test is shown in Fig.~\ref{fig:scam-vs-fullcam} (a).
The plot uses \eq{eq:gamma1sol} to express the dependence of the mass ratio $\gamma_{1}$ on the secondary star distance (via $\alpha_{2}$).
It shows two shaded regions that overlap each other partially.
The darker shade represents the solutions according to the simplified CAM, while the lighter shade are the results from the full CAM.
Apparently the simplified CAM shows the same qualitative behaviour and follows the full solutions closely, but with \edit1{an offset} in vertical direction (note the logarithmic scale of the $y$-axis).
This deviation does not show any dependence on the secondary star distance (i.e. $\alpha_{2}$) or on the secondary star's eccentricity (not shown).
We have to bear in mind that \eq{eq:gamma1sol} involves \edit1{strong approximations with respect to} the distance ratios $\alpha_{1}$ and $\alpha_{2}$.
Accordingly, the major effect causing the visible deviation is the magnitude of $\alpha_{1}$ (fixing $a_{GP} = 5.2$~au).
\edit1{The deviation becomes larger with increasing distance ratio $\alpha_{1}$, i.e. when the GP is near to the HZ and its perturbations are stronger.}

Figure~\ref{fig:scam-vs-fullcam} (b) shows another comparison of the accuracy.
In this case we use \eq{eq:alpha2sol0} to trace the distance ratio $\alpha_{2}$ as a function of secondary star mass in the range $0.08 \le m_{B} \le 2.1 \; M_{\odot}$.
We set the giant planet to 5~au and always assign an eccentricity of $e_{B} = 0.25$ to the secondary.
The results demonstrate that the simplified CAM is able to reliably describe the dynamics, albeit with a small deviation.
It is visible from the figure that the simplified CAM retrieves a narrower region in the parameter space than the full CAM, but the deviation is only minor especially at the lower border.
Similar to the first test case the parameter $\alpha_{1}$ plays a major role for the magnitude of the deviation: the larger $\alpha_{1}$ the larger is the deviation.

We can conclude that the simplified CAM is able to \edit1{reliably describe the dynamics}, provided that the giant planet does not approach the HZ too closely.
\edit1{Its limitations are clearly connected to the parameter $\alpha_{1} = a_{TP} / a_{GP}$ that should be $\alpha_{1} < 1/3$ (in particular $< 0.1$) in order to avoid the effect of MMR that would invalidate a purely secular model.}
The approximate solutions can thus serve to quickly estimate the variation in \edit1{any of the five system parameters}, and to scan a large parameter space for a first overview \edit1{on possible secular perturbations in the HZ.}


\subsection{Application to exoplanet systems of binary stars} \label{sec:result:apply}

Before continuing to present the model outcomes applied to synthetic systems, we need to be more explicit about the choice of system parameters.
As mentioned above, we consider only perturbers external to the HZ \edit1{of a G-type main sequence star}.
This puts a constraint on the location of the giant planet \edit1{to orbit beyond about $2 - 3$~au depending on its eccentricity}.
Thus, in a first step, we need to get an overview of suitable exoplanet candidates.


\subsubsection{Selection process} \label{sec:result:select}

\begin{figure}
    \centering
    \includegraphics[width=0.5\textwidth]{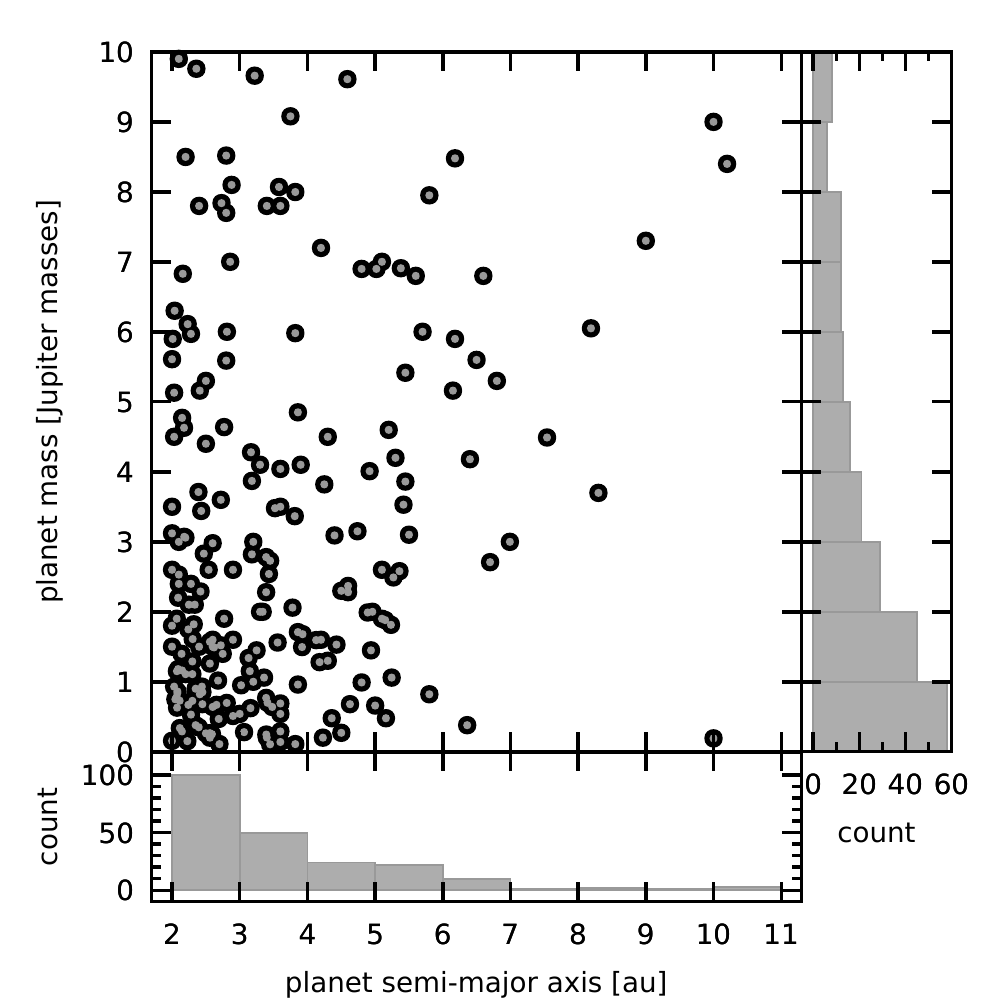}
    \caption{Sample of 220 confirmed exoplanets from the Exoplanet.eu database. The selected planets are restricted to masses in the interval from 0.1 to 10 Jupiter masses, and semi-major axes from 2 to 11 au. The histograms show the distributions of the respective parameter.}
    \label{fig:exoplanets}
\end{figure}

We selected from \edit1{various online catalogs\footnote{
For instance, The Extrasolar Planets Encyclopaedia of \citet{Sch2011} at \url{https://exoplanet.eu/catalog/}, or the NASA Exoplanet Archive \url{https://exoplanetarchive.ipac.caltech.edu/index.html}.}} all confirmed exoplanets that fulfill the following two criteria: (1) masses between $0.1-10$ Jupiter masses ($M_J$), and (2) semi-major axes between 2 and roughly 10 au, irrespective of planet or stellar multiplicity.
This gives a total of 220 exoplanets in our sample.
The choice of parameters is motivated first by the evident incompleteness of planet detections at larger distances, and second by the survey of \citet{Bry2016}.
From their combined radial velocity -- imaging survey those authors concluded that the frequency of giant planets peaks between 3 and 10~au for the mass range up to 20~$M_{J}$.

Figure~\ref{fig:exoplanets} shows the scatter plot and respective parameter distributions of these exoplanets.
The histogram on the right shows that the median (minimum) mass is around 2.5~$M_J$, with an extended tail reaching up to the selection limit.
The distribution of semi-major axes in the bottom histogram is more concentrated to lower values, with about 90~\% of the sample having values of $a_{GP} \le 6$~au.
This is a consequence of the detection by mainly radial velocity surveys that favour exoplanets with shorter orbital periods \citep{Cum2008, Per2011}.

This selection of giant planet properties has implications on the secondary star parameters.
We restrict ourselves to masses in the interval $0.4 \le m_{B} \le 1.3$~$M_{\odot}$ (roughly spectral types M2V -- F5V), and a lower limit on the stellar separation of $a_{B} \ge 20$~au.
The eccentricities can vary between $0 \le e_{B} \le 0.6$ according to the fit limitation in the model of \citet{And2017}.


\subsubsection{Binary star parameters} \label{sec:result:binpar}

As we can appreciate from \eq{eq:freqeqmain} the right-hand side has a non-linear dependence on $\gamma_{2}$, $\alpha_{2}$, and  $e_{B}$.
This means that the location of the SR will strongly depend on the secondary star's parameters.
We will first investigate the qualitative behaviour of the solutions.

In a former study, \citet{Pil2016} have investigated the system HD~41004~AB as a proxy for binary stars with circumstellar exoplanets.
They have demonstrated that the positions of SR are connected to the orbital parameters of the secondary star, especially to its distance.
Accordingly, they have investigated the qualitative behavior of SR location for a shift in $a_{B}$ (for $a_{GP} = $~const.).
Here, our purpose is different: We constrain the SR to lie in the HZ and show how the system parameters (masses and distances of GP and secondary) need be coupled in order to maintain that situation.

From Fig.~\ref{fig:motiv} we can observe the correlation between $a_{B}$ and $a_{GP}$ that results in the diagonal stripe.
This means that for each value of $a_{B}$ (for fixed $a_{TP}$) there is a unique solution to \eq{eq:freqequal} that yields a distance $a_{GP}$ causing an SR in the HZ.
The figure contains another important detail: When increasing the secondary's distance from the host star (for fixed GP distance along the horizontal lines, from left to right) we can observe that the SR moves from outside in.
This means that first the outer part of the HZ (upper edge of the stripe) would be affected, then some intermediate region of the HZ, and finally the inner border (lower edge).
Increasing the distance even more would then remove the SR from the HZ and would place it closer to the host star.
Such an inverse relation between $a_{B}$ and the SR location is not immediately obvious from \eq{eq:freqeqmain}, but we can understand it in terms of $\alpha_{2}$.
The variable $\alpha_{2}$ is simply the ratio $a_{GP} / a_{B}$, hence it is inversely proportional to the secondary's distance.
In case that the other parameters are held fixed (as in Fig.~\ref{fig:motiv}), an increase of $a_{B}$ leads to a decrease of $g_{GP}$ because of the term $\propto \alpha_{2}^{3}$, cf. \eq{eq:heppenh}.
Conversely, the frequency $g_{TP}$ also has to decrease to satisfy \eq{eq:freqeqmain}.
The balance is restored through the term $\propto \gamma_{1} \alpha_{1}^{2}$ (that is a monotonic function and generally dominates over the second term) via the decrease of $\alpha_{1}$, i.e. the resonance location moves away from the giant planet and closer to the host star.
A physical interpretation of this behaviour is that a more distant perturbing secondary star reduces the rate of angular momentum exchange with the giant planet -- its secular frequency $g_{GP}$ decreases.
Similarly, test planets have lower orbital precession frequencies the further away they are from the GP (which is their main perturber), such that necessarily the resonance location moves away from the GP.

At the left edge of the Fig.\ref{fig:motiv}, the red line denotes the critical distance, $\acrit$, of the GP according to the fits of \citet{Hol1999} and \citet{Pil2002}.
This line separates the dynamically unstable area (hatched, to the left) from the region of orbital stability.
The indicated example with $e_{B} = 0$ is an optimistic best-case scenario, typically the secondary star has a non-zero eccentricity.
Furthermore, the results of \citet{Pil2002} suggest that the hatched region would also be more extended if we accounted for the GP's elliptic motion.

One of the main observables of binary star systems is the mass of the secondary star.
This parameter is of utmost importance to dynamical studies, because it contributes significantly to the size of the host star's region of orbital stability and determines the strength of gravitational perturbations.
In the following example we show the variation of the SR's location with the secondary star's mass.

\begin{figure}
    \centering
    \includegraphics[width=0.5\textwidth,page=1]{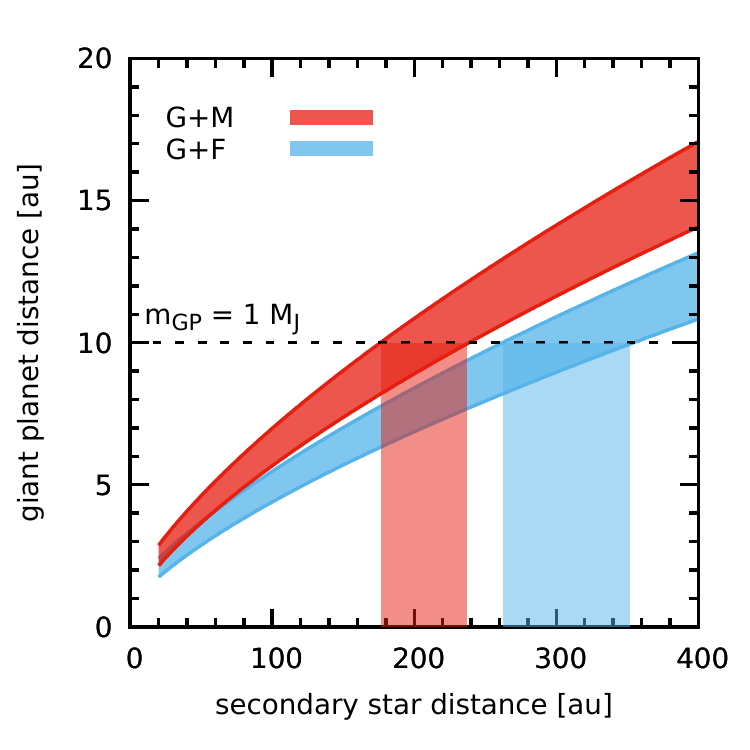}
    \caption{\edit1{Resonance locations as a function of the secondary star's mass.} These two cases show a a G+M binary (top curve), and a G+F binary (bottom curve). The corresponding rectangles show secondary star distances (for a GP at 10~au) that perturb the HZ.}
    \label{fig:secmass}
\end{figure}

Figure~\ref{fig:secmass} presents two binary stars in the same parameter space like before.
The G+M (upper curve; $m_{B} = 0.4 \, M_{\odot}$) and G+F (lower curve; $m_{B} = 1.3 \, M_{\odot}$) systems relate to edge cases with lowest and highest secondary star masses, respectively.
All other mass combinations (e.g. G+G, G+K) fall in between these two curves.
In this example there is a giant planet with one Jupiter mass at 10~au (dashed horizontal line).
This planet induces an SR in the HZ for secondary star distances of $180 - 240$~au in the G+M case, and $260 - 350$~au in the G+F case.
There is a clear trend with secondary mass: the curve becomes flatter with increasing mass, and the absolute width (in terms of $a_{B}$) of the SR generating region broadens.

We can thus derive a general picture from Figs.~\ref{fig:motiv} and \ref{fig:secmass}: (1) there is a positive correlation between giant planet and secondary star distance for fixed secondary star mass; and (2) the secondary's mass directly relates to the interval of stellar distances generating SR for fixed giant planet distance.
This means that a higher mass secondary at a distance of several hundreds of astronomical units might still perturb the GP in such a way as to induce an SR in the HZ.


\subsubsection{Binary star and planet parameters} \label{sec:result:planpar}

Until now some planet-related parameters were held constant or ignored altogether.
Next we will consider also the variation of these planet parameters and investigate their influence on the results.

\begin{figure}
    \centering
    \includegraphics[width=0.5\textwidth]{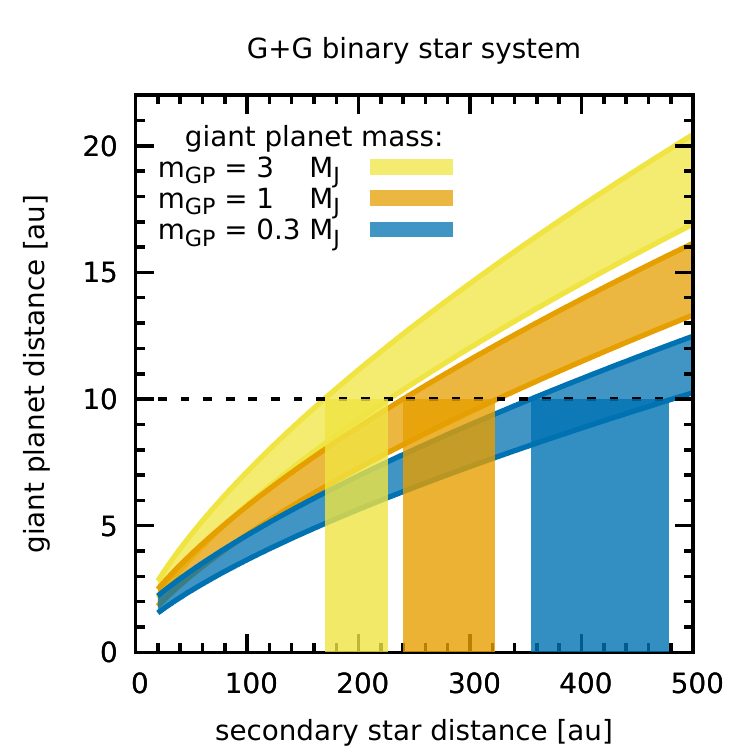}
    \caption{\edit1{Resonance locations as a function of the GP's mass.} The plot shows a G+G binary star system for three different GP masses (in Jupiter masses, $M_{J}$). The GP is always at a distance of 10~au (dashed line).}
    \label{fig:gpmass}
\end{figure}

We have already demonstrated in Fig.~\ref{fig:secmass} that there is a correlation of stellar mass and resonance location.
A similar correlation also exists for the planet mass, although with the inverse effect as Fig.~\ref{fig:gpmass} shows.
This figure displays the same parameter space ($a_{B}, a_{GP}$) like the ones before.
In this example we have a couple of G-type stars (each with $1 \, M_{\odot}$) with a planet located at 10~au whose mass is a free parameter.
The top curve is for a planet with 3 times the mass of Jupiter ($M_{J}$), the middle curve for 1~$M_J$, and the bottom curve corresponds to a Saturn-mass planet.
As the GP's mass increases (from bottom to top) the curve becomes steeper and the resonance location shifts to the left towards smaller secondary star distances.
The consequences of this behaviour for different planet masses are twofold: (1) a negative feedback for more massive planets, and (2) a feedback through coupling of planetary mass and eccentricity.
To clarify both issues we will compare two systems for equal planetary distances but with different masses.

It is intuitively clear that a more massive planet is a more efficient perturber of the HZ area than its lower mass counterpart.
However, at the same time, a higher mass planet also requires smaller stellar distances \edit1{to force the SR into the HZ}, as it is visible from the figure.
These two facts entail stronger perturbations in the HZ due to larger planet mass and closer secondary star.
Hence, the negative feedback consists in an increased perturbation strength for a higher mass planet in combination with a closer secondary star.

The second feedback effect concerns the opposite effect.
With decreasing planet mass the perturbing secondary star has to shift outwards to a distance of several hundred au to place an SR into the HZ.
Such extended systems have a weaker dynamical coupling than more compact ones, hence one would not expect strong perturbations in the HZ.
However, the giant planet's eccentricity plays a crucial role and compensates for the weaker coupling to the secondary star, see Fig.~\ref{fig:emaxhz} (b).
In a wide binary system the second star is more susceptible to external perturbations e.g. from passing stars (see section \ref{sec:discuss:flyby}).
Such perturbations can act to increase the secondary star's eccentricity.
The giant planet's forced eccentricity is approximately $\propto e_{B} / (1 - e_{B}^{2})$ \citep[see][]{And2017}, so it increases with larger $e_B$.
Additionally, in former studies we have found that observed exoplanets have higher eccentricities in general than what their forced eccentricity would suggest \citep[see][Table~1]{Baz2017}.
Although the secondary star must be farther away for lower mass planets it can effectively induce a higher planetary eccentricity.
In this way the increased eccentricity can balance the magnitude of perturbations that otherwise would decrease with smaller planetary mass.

\begin{figure}
    \centering
    \includegraphics[width=0.5\textwidth]{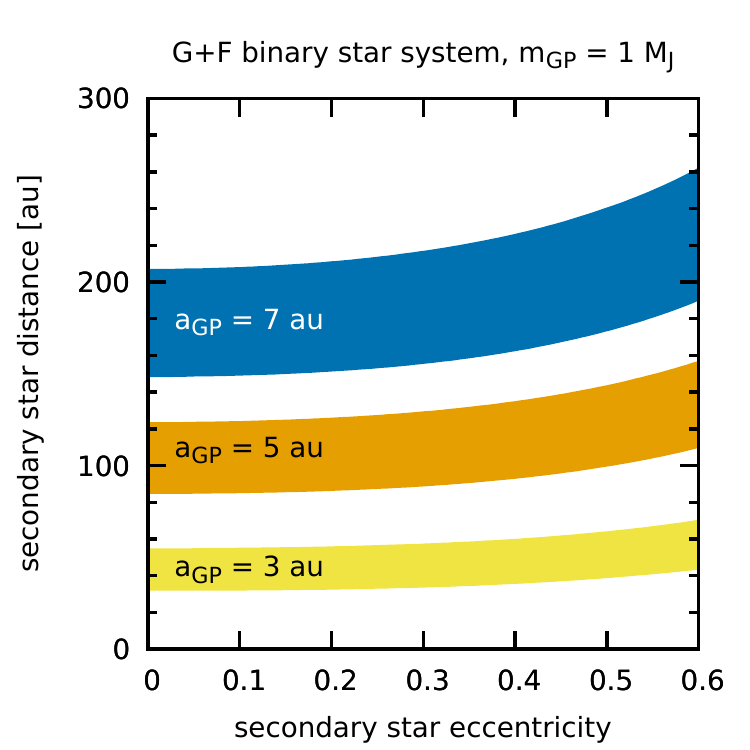}
    \caption{\edit1{Resonance locations as a function of the secondary star eccentricity and giant planet distance.} The perturbed zone falls to different locations depending on the giant planet's distance. This zone also shifts depending on the secondary star's eccentricity.}
    \label{fig:sececc}
\end{figure}

The eccentricity does not only play a role for the planet, but -- more importantly -- also for the stellar orbit.
Figure~\ref{fig:sececc} shows a different cut through the ($e_{B}, a_{B}$) parameter space.
Three cases present the resonance location as a function of the secondary's eccentricity and the planet's location (here the GP has one Jupiter mass).
When placing the GP to 3, 5, or 7~au (from bottom to top), respectively, the SR is induced by secondaries in a wide range of distances ($y$-axis).
The SR location also depends on the secondary star eccentricity: as $e_{B}$ grows the SR moves outwards with respect to the circular case.
We can identify the cause for this variation with the terms $\propto (1 - e_{B}^{2})^{-3/2}$ in eq.~(\ref{eq:freqeqmain}).
These two terms are responsible for the weak deviation from the circular value at low $e_{B}$, and the increasingly larger difference at high $e_{B}$.

A striking characteristic of this figure pertains to the large shifts in $a_{B}$ for a minor change of $a_{GP}$.
For instance, the SR is displaced from about 50~au (for $a_{GP} = 3$~au) to roughly 200~au (for the GP at 7~au).
These results indicate potentially dramatic changes of the system's secular architecture by slight shifts in $a_{GP}$ (see section \ref{sec:discuss:flyby}).


\subsection{Resonance generating distance of giant planet} \label{sec:result:relation}

In Figures \ref{fig:motiv}, \ref{fig:secmass}, and \ref{fig:gpmass} we can observe that the distances of GP and secondary star show a positive correlation, but also that this relation is basically nonlinear.
It is very important to understand the occurrence of SR in the HZ as a function of the GP's semi-major axis, $a_{GP}$, for actual applications.
Hence, we need to understand the relation between $a_{GP}$ and $a_{B}$ as a function of the other parameters.
For this we cannot use the solutions presented in Appendix \ref{sec:app:scam:alpha2sol}, because they express $\alpha_{2}$ in terms of $\alpha_{1}$ which itself contains $a_{GP}$, so those equations only give implicit relations.
Conversely, in the following we will derive an explicit relation on the basis of \eq{eq:freqeqapprox}.

We start from the simplified CAM in the limit $\delta \rightarrow 0$ to get rid of the additional dependence on $\alpha_{2}$.
Recall the definitions $\alpha_{1} = a_{TP} / a_{GP}$ and $\alpha_{2} = a_{GP} / a_{B}$, and that the constant $a_{TP}$ is either $\ahzin$ or $\ahzout$ for the inner or outer HZ border, respectively.
After inserting the corresponding variables into \eq{eq:freqeqapprox}, \edit1{we obtain a cubic equation for $z = a_{GP}^{3/2}$.}
The general solution is thus given by
\begin{equation} \label{eq:agpsol}
    \alpha_{2} = 3^{-2/3} \left( \frac{a_{TP}}{a_{B}} \right) \left( \frac{(u + u^{-1} - 1)^{2}}{1 + \gamma_{1}} \right)^{1/3},
\end{equation}
with the abbreviations
\[
    u = \left( -2 c_{2}^{3} - 27 c_{0} c_{3}^{2} + 3 \sqrt{3} \sqrt{4 c_{0} c_{2}^{3} c_{3}^{2} + 27 c_{0}^{2} c_{3}^{4}} \right)^{1/3} / \left( 2^{1/3} c_{2} \right),
\]
and $c_{0} = - \gamma_{1} a_{B}^{3}$, $c_{2} = - \gamma_{2} \epsilon_{B}$, $c_{3} = (1 + \gamma_{1})^{1/2} \gamma_{2} \epsilon_{B} a_{TP}^{-3/2}$.

An important property of solution (\ref{eq:agpsol}) is that the expression $\gamma_{2} \epsilon_{B}$ appears only in this form, i.e. these parameters are never separated from each other.
Furthermore, we could expand the denominator into a power series $(1 + \gamma_{1})^{-1/3} \approx 1 - \gamma_{1} / 3 + O(\gamma_{1}^{2})$ since $\gamma_{1} = m_{GP} / m_{A} \ll 1$.
Note that the ratio $a_{TP} / a_{B}$ also plays a role in $u$ via $c_{0} c_{3}$.

Equation (\ref{eq:agpsol}) is useful to calculate the orbital distances of GP that would generate secular perturbations in the HZ for any given binary star system.
Once the corresponding value of $\alpha_2$ is known, we can determine the GP location from $a_{GP} = \alpha_{2} a_{B}$ and construct parametric plots like in Fig.~\ref{fig:motiv}.

Tables~\ref{tab:dataGF} -- \ref{tab:dataGM} in Appendix \ref{sec:app:tabs} serve to assess the accuracy of \eq{eq:agpsol}.
\edit1{%
Each table lists the critical values $\alpha_{2,\text{IHZ}}$ and $\alpha_{2,\text{OHZ}}$ (for the inner and outer HZ border, respectively) for a given binary star system.
These solutions of \eq{eq:freqeqmain} determine the location of the GP, $a_{GP}$, as a function of the secondary star's semi-major axis, $a_{B}$, such that an SR affects some part of the HZ.
The main parameters are the GP's mass (in units of Jupiter's mass), and the secondary star's eccentricity $e_{B}$.
}
The general trends of (1) decreasing $\alpha_2$ with increasing $a_{B}$ (i.e. the negative curvature, also visible in Fig.~\ref{fig:motiv}), and (2) the slight shift in $\alpha_{2}$ with increasing $e_{B}$, like in Fig.~\ref{fig:sececc}, are clearly discernible from the data, too.

\begin{figure*}
    \centering
    \includegraphics[width=0.8\textwidth]{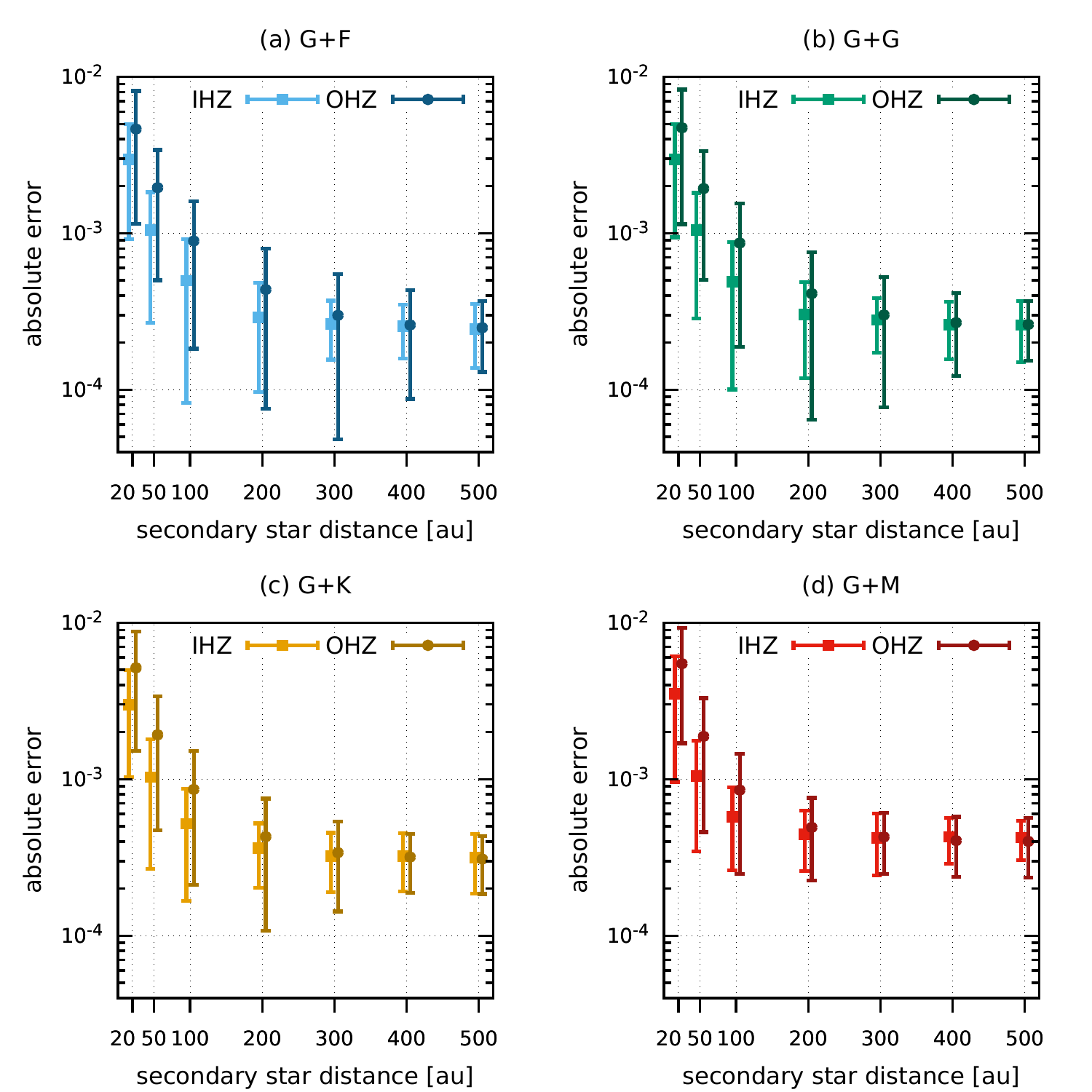}
    \caption{Comparison of absolute deviations for $\alpha_{2}$ between the model from \eq{eq:agpsol} and numerical solutions based on \eq{eq:freqeqmain} (cf. tables in Appendix \ref{sec:app:tabs}). Each pair of error bars represents averaged deviations for a certain value of $a_B$. Symbols with squares (lighter colour) represent the deviations for the IHZ, while bullets (darker colour) mark the deviations for the OHZ. Panels (a) to (d) give details for different types of binary star systems.}
    \label{fig:fitdiff}
\end{figure*}

The numbers in the tables have been used as reference values for comparison with predictions  by employing the model in \eq{eq:agpsol}.
Figure~\ref{fig:fitdiff} shows the absolute deviations $| \Delta \alpha_{2} |$ between the model and the reference values from the tables.
For each distance bin in $a_{B}$ we computed the average error and its standard deviation (1-$\sigma$) taken over all different combinations of $m_{GP}$ and $e_{B}$.
The figure shows that for $a_{B} \ge 100$~au the error is less than $10^{-3}$, i.e. the location of the GP that generates an SR can be determined with a typical accuracy of $< 0.1$~au.
With smaller secondary star distance the errors are larger though, because this is a direct consequence of neglecting higher order terms in the expansion of the Laplace coefficients when deriving the simplified CAM.
A closer look at the figures also reveals the trend that the errors grow steadily from G+F to G+M stars (left to right).
This behaviour can be understood  by the dependence of the SR region on the secondary's mass as mentioned in connection with Fig.~\ref{fig:secmass}: a more distant secondary must be more massive to generate the same effect as a more proximal lower-mass star.
Thus, M-type companion stars are naturally closer to the GP, and so the ratio $\alpha_2$ is larger, which leads to somewhat higher deviations when using \eq{eq:agpsol}.




\section{Discussion} \label{sec:discuss}

Results from earlier investigations of binary star systems assumed that well-separated binaries (beyond roughly 200~au) can be simply treated as single star systems \citep[e.g.][]{Des2007}.
The new results presented in the previous section shed a different light on those assumptions, they demonstrate that secular perturbations do affect the HZ for a wide range of binary system orbital and physical parameters.
In the following sections we will discuss some of the implications of those resonances on the dynamical evolution of test planets in the HZ.


\subsection{Effect of stellar fly-bys} \label{sec:discuss:flyby}

Passing stars can affect planetary systems severely and can lead to direct or indirect ejections \citep[see review in][]{Dav2014}.
A subtle effect relates to the delayed instability of planetary systems of wide binary stars induced by external perturbations, as discussed by \citet{Kaib2013}.
They explain the apparent overabundance of eccentric (circumstellar) exoplanets in wide binary systems by the action of galactic tides that lead to an increase of the secondary star's eccentricity.
At a certain point its eccentricity is large enough for the periastron distance to drop below a critical value, below which the star triggers the instability of the exoplanetary system.

\citet{Cor2017} reiterate on this topic to disentangle the individual contributions of two types of external perturbations: (1) the galactic potential, and (2) stellar encounters via fly-bys.
They find that stellar passages have a destabilizing effect, because 40~\% of their test binary systems are disrupted, and this value is (nearly) independent of the systems's initial eccentricity \citep[Fig.~2]{Cor2017}.
This means that a stellar fly-by, by itself, is effective in modifying the secondary's eccentricity.

These results are verified by \citet{Ban2019}, who study the dynamical effects of a sequence of stellar encounters between a binary star system and a passing star.
Their results show that the stellar eccentricity has a maximum variance that reaches up to 0.45 for an initial stellar separation of $\ge 150$~au.
Hence, stellar fly-bys over an extended period of time can significantly modify the companion star's orbital eccentricity.

Returning to Fig.~\ref{fig:sececc}, we can put these findings about stellar fly-bys into context with the results shown in that figure.
Let us assume a binary system with an architecture such that initially the HZ is free of SR, i.e. the system would be located somewhere in the white area above one of the coloured zones.
The combined effect of stellar fly-bys then can lead to a significant change (increase) of the secondary star's eccentricity, perhaps in such a way that it enters one of the coloured bands.
Consequently, an exoplanetary system with a dynamically calm HZ can transition into one that is affected by strong secular perturbations.
This could turn a planet that has always had habitable conditions into a potentially uninhabitabe one, depending on how much its eccentricity increases by the perturbations.
The next section considers how to exactly quantify the effect of perturbations.


\subsection{Related works} \label{sec:discuss:other}

The scope of the presented model is not limited to identify and trace SR throughout the HZ of binary star systems.
\edit1{%
In an alternative embodiment of this method we could prescribe any interval of interest, not just the HZ.
For instance, an application could be to check whether or not an SR occurs at or around the location of a detected or candidate exoplanet in a multi-planet system.
}
This can be achieved by selecting for $a_{TP}$ an interval of distances that is different from the HZ range $\ahzin \le a_{TP} \le \ahzout$.

In the present study a major premise is that all bodies are coplanar or nearly so.
As this is not generally the case, we have to consider the effect of mutual inclinations of the planets and the outer perturber.
\citet{Sal2009} have investigated the dynamics and stability of initially coplanar two-planet systems in tight binaries with an inclined stellar companion.
They identified a narrow zone for the second planet that suppresses the Kozai-Lidov effect on an inner hot-Jupiter and helps to stabilize the system.
\citet{Den2019} improve on these former results by studying basically the same situation, but they derive an analytic stability criterion that takes into account the perturber's separation and eccentricity.
Their study might be viewed as an extension of the current work to more degrees of freedom by adding an inclination to the system.
They include the additional perturbations due to the Kozai-Lidov mechanism, however they only consider planetary mass perturbers in lieu of stellar masses like in our work.
\edit1{%
An extension of the current model could follow the lines of \citet{Li2017} who studied SR for prograde and retrograde orbits.
They constructed a semianalytical model for the restricted four-body problem that allows to describe both types of motion with a unified formalism following a suitable change of reference system.
In this model resonant angles consist of linear combinations of angular differences of apsidal and nodal precession frequencies.
Although the dynamical model of \citet{Li2017} is similar to the one used in this paper (see section \ref{sec:method:dynmod}), they studied planetary satellites in very close orbits whereas we focus on rather hierarchical systems with period ratios $P_{\text{out}}/P_{\text{in}} > 2$.
}

Furthermore, in analogy to the current work, \citet{Yel2018} have studied the influence of SR on a debris disc for the case of interior planetary perturbers.
In particular, they mapped how the locations of SR depend on the masses, semi-major axes, and eccentricities of the planets.

The work of \citet{Mar2016} also contains certain aspects from the current study.
When they investigate the stability of multiplanet systems in binaries by frequency map analysis, they identify cases when the two planets are not long-term stable although both of them are located well inside the critical stability border.
They attribute this to the effect of the binary's eccentricity and establish an empirical relationship for $(a_{B}, e_{B})$ to predict a critical value $a_{B,\text{crit}}$ below which the planets become unstable.
For computational reasons they restricted their parameter range to $10 < a_{B} < 100$~au and to only three different stellar mass ratios.
The CAM can be regarded as a generalization of their empirical relation that takes into account a much larger parameter space and hence covers a broader variety of use-cases.
Our CAM model could also offer an explanation for the bifurcated behaviour of eccentricity excitations of exoplanets under binary perturbations that \citet{Tak2008} describe in their Fig.~14.
They noted that Kozai-cycles can be either suppressed or fully active in the same parameter space region for a small change in the planet's parameters.

In \citet{Pu2018} the authors focus on the same topic of a two-planet system subject to the perturbations of an external perturber.
Their analysis aims to derive analytical expressions for the eccentricities of the inner planetary system as a function of the perturber's orbital elements and mass.
A main result is that a weak secular coupling between the planets (as is the case in hierarchical systems) leads to a higher susceptibility for eccentricity excitations, especially if the innermost planet is the least massive one, similar to what we have demonstrated in section \ref{sec:method:emax}.




\section{Summary} \label{sec:summary}

In this study, we investigate circumstellar planetary systems of binary stars \edit1{(with coplanar and prograde motion)} and the conditions that lead to secular perturbations in the habitable zone of the host star.
Our main aim is to identify zones in the five-dimensional parameter space (orbital and physical parameters of the massive bodies, cf. Fig.~\ref{fig:dynmodel}) that generate \edit1{dynamically} unfavourable conditions for potentially habitable planets, \edit1{such as elevated orbital eccentricities.}
This contribution helps to identify such systems in large-scale surveys and to exclude them from in-depth observations.

We apply the restricted four-body problem as dynamical model and construct a Combined Analytical Model (CAM) to systematically map the properties of different synthetic systems (see section \ref{sec:method:cam}).
Subsequently, we introduce additional simplifications to the CAM (section \ref{sec:method:simpcam}) and derive explicit analytical formulas to express the resonance locations in each system parameter as a function of the others (see Appendix \ref{sec:app:scam}).
The CAM is assessed in comparison to fully numerical results which shows that it has a median relative error less than 3~\% (cf. section \ref{sec:result:modelacc}).

Our results are based on the properties of a sample of detected exoplanetary systems from different online databases.
Secular perturbations in the HZ appear even for hierarchical wide binary systems, where stellar separations are on the order of several hundred astronomical units.
We give details on the parameter combinations that generate SR in the HZ in the form of two-dimensional sections through the parameter space (section \ref{sec:result:apply}).
Additionally, we derive analytical expressions to calculate the set of unfavourable giant planet positions for any given mass, semi-major axis, and eccentricity of the secondary star (in section \ref{sec:result:relation}).

Based on these results we introduce the online tool \emph{SHaDoS} (Appendix \ref{sec:app:tool}) that implements the CAM and provides a simple user interface to reproduce the results presented in this study.



\acknowledgments

AB and EPL are grateful for the continuous support from Austrian Science Fund (FWF) project S11608-N16, part of the National Research Network ``Pathways to habitability'' (PatH).
We wish to thank our colleagues D. Bancelin, N. Georgakarakos, and C. Lhotka for inspiring discussion in the course of preparing this manuscript.
The authors thank an anonymous reviewer whose comments and suggestions helped to improve the manuscript.





\appendix

\section{Online Tool} \label{sec:app:tool}

We provide the online tool \emph{SHaDoS}\footnote{This is an acronym for \emph{S}ecular perturbations in \emph{Ha}bitable zones of \emph{Do}uble \emph{S}tars.
The tool is accessible at \url{https://www.univie.ac.at/adg/shados/index.html}} and allows to reproduce the results presented in this article.
It implements the CAM model introduced in section~\ref{sec:method} and solves \eq{eq:freqequal} for any given set of parameters.
In the following we describe the main features of this tool.

The application implements an object-oriented approach with a step-by-step process for user input.
There are four consecutive steps tied to the main objects of interest: (1) the host star, (2) the perturbing planet beyond the HZ, (3) the distant secondary star, and (4) the parameter space.
In each of these substeps the application queries for some system parameter, refer to Fig.~\ref{fig:dynmodel} for an overview:
\begin{enumerate}
    \item The relevant host star parameters are mass, and spectral type.
    Four different spectral types are preset, namely F, G, K, and M-type main-sequence stars with their respective stellar luminosity and effective temperature.
    Users can manually enter any other stellar characteristics, but effective temperatures are limited to the interval $2600 \le T_{\text{eff}} \le 7200$~K.
    The extent of the HZ is then computed based on this input.

    \item Giant planet parameters include mass (in units of Jupiter's mass), distance to the host star (semi-major axis), and orbital eccentricity.
    The eccentricity is restricted to elliptic orbits, and the distance must be larger than the outer HZ border $\ahzout$.

    \item The input for the secondary star covers its mass, distance, and eccentricity.
    Eccentricity is restricted to the interval $0 \le e_{B} \le 0.6$ due to limitations in the analytical model (cf. section~\ref{sec:method:dynmod}).

    \item The parameter space consists of a two-dimensional grid, where a primary parameter must be selected as the independent variable, and a secondary parameter is defined as the sought dependent variable.
    Available parameters are any of those depicted in Fig.~\ref{fig:dynmodel} except for the already fixed host star mass $m_{A}$.
    Additionally, the minimum and maximum value of the independent variable must be entered, as well as the number of discretization steps.
\end{enumerate}

After having finished all of these steps the online tool performs the calculations on-the-fly and finally displays the results.
There is an option to save the resulting plot as an image in the widely supported PNG format.
Any comments, suggestions, or bug reports are welcome and should be addressed to the first author.


\section{Solutions for simplified CAM} \label{sec:app:scam}

Here we provide analytical solutions of the simplified CAM from \eq{eq:freqeqapprox}.
The main goal is to express each of the parameters $(\gamma_{1}, \gamma_{2}, \alpha_{1}, \alpha_{2}, \epsilon_{B})$ in turn as a function of the other four variables.


\subsection[Solutions for gamma1]{Solutions for $\gamma_1$} \label{sec:app:scam:gamma1sol}

There exist two real-valued solutions for $\gamma_{1}$,
\begin{equation} \label{eq:gamma1sol}
    \gamma_{1, \pm} = \eta \left[ -2 \alpha_{1}^{3} \pm (1 - \delta) \left( \eta (1 - \delta) + \sqrt{4 \alpha_{1}^{3} (1 - \eta) + \eta^{2} (1 - \delta)^{2}} \right) \right] / 2 \alpha_{1}^{3},
\end{equation}
where $\eta = \gamma_{2} \alpha_{2}^{3} \epsilon_{B}$, and $\delta$ is given by \eq{eq:andeggcorr}.
The square-root is definitely positive if $1 - \eta \ge 0$, i.e. if $\gamma_{2} \alpha_{2}^{3} \epsilon_{B} \le 1$.
This puts a coupled constraint on these three parameters, because it holds strictly that $0 \le \alpha_{2} < 1$, and $\epsilon_{B} \ge 1$, while $\gamma_{2}$ can be either smaller or larger than 1.
Moreover, we can discard $\gamma_{1,-} < 0$ as an unphysical solution, because the mass-ratio cannot be negative.
The remaining $\gamma_{1,+}$ is the preferred solution to \eq{eq:freqeqapprox}.


\subsection[Solution for alpha1]{Solution for $\alpha_1$} \label{sec:app:scam:alpha1sol}

Apart from the trivial solution $\alpha_{1} = 0$, we find
\begin{equation} \label{eq:alpha1sol}
    \alpha_{1} = (1 + \gamma_{1})^{1/3} \left( \frac{\eta (1 - \delta)}{\gamma_{1} + \eta} \right)^{2/3}.
\end{equation}
Note that additionally there exists also a pair of complex conjugated solutions for $\alpha_{1}$, but again these are not of physical interest.
We caution that this functional form is only a first approximation to the `true' solution of the (simplified) CAM.
The formula above will give satisfactory results only if $\alpha_{1} \ll 1$, i.e. typically for $\alpha_{1} < 0.1$, because among all parameters $\alpha_{1}$ is most sensitive to the assumptions made about the magnitudes of the distance ratios $\alpha_{i}$ ($i = 1,2$).
In many cases $\alpha_{1}$ will be larger than this threshold, compare the histogram of semi-major axes in Fig.~\ref{fig:exoplanets}.


\subsection[Solutions for gamma2]{Solutions for $\gamma_2$} \label{sec:app:scam:gamma2sol}

While it is straightforward to obtain the solutions for $\gamma_{1}$ and $\alpha_{1}$, there is a caveat regarding the other parameters.
The problem is the explicit polynomial dependence of $\delta$ on the parameters $\gamma_{2}$, $\alpha_{2}$, and $e_{B}$ (cf. \eq{eq:andeggcorr}), which leads to rather complicated equations.
One way to circumvent this issue is to linearize the equation in $\gamma_2$ by the approximation $\delta \rightarrow 0$.
This is equivalent to a fall-back to the Heppenheimer formula \eq{eq:heppenh} and hence gives an approximation of first-order in masses.
The unique linearized solution for $\gamma_{2}$ reads
\begin{equation} \label{eq:gamma2sol}
    \gamma_{2}^{(0)} = \frac{\gamma_{1} \alpha_{1}^{3/2}}{\alpha_{2}^{3} \epsilon_{B} \left( (1 + \gamma_{1})^{1/2} - \alpha_{1}^{3/2} \right)},
\end{equation}
where the superscript 0 denotes the approximative nature of this solution in the limit $\delta \rightarrow 0$.
This solution is defined for any combination of parameters, since $(1 + \gamma_{1})^{1/2} \ge 1$ and $0 \le \alpha_{1}^{3/2} < 1$, so the denominator is always positive.

When considering the full equation for $\gamma_{2}$ we only sketch the steps to find the solutions.
From \citet{And2017} we observe that the correction term is $\delta(\gamma_{2}) \propto F(\gamma_{2}^{1/2}) + G(\gamma_{2}^{1}) + H(\gamma_{2}^{2})$, where the functions $F,G,H$ also depend on $\alpha_{2}$ and $e_B$.
Using the substitution $z = \gamma_{2}^{1/2}$ this results in a polynomial equation of degree 6 in $z$, for which the roots can be found with the Newton-Raphson method with initial guess $\gamma_{2}^{(0)}$.


\subsection[Solution for alpha2]{Solution for $\alpha_2$} \label{sec:app:scam:alpha2sol}

On inserting the secular frequency correction term $\delta$ from \eq{eq:andeggcorr} into \eq{eq:freqeqapprox} we face the problem that it would introduce fractional powers of $\alpha_2$.
Hence, we restrict ourselves to the same approximation as before,$\delta \rightarrow 0$, such that no additional terms proportional to $\alpha_2$ arise.
The resulting equation possesses the solution
\begin{equation} \label{eq:alpha2sol0}
    \alpha_{2}^{(0)} = \left( \frac{\gamma_{1} \alpha_{1}^{3/2}}{\gamma_{2} \epsilon_{B} ((1 + \gamma_{1})^{1/2} - \alpha_{1}^{3/2})} \right)^{1/3}.
\end{equation}
Note that it resembles \eq{eq:gamma2sol}, but the roles of $\gamma_{2}$ and $\alpha_{2}$ are exchanged.
A better approximation is found if we kept in $\delta$ the term $\propto \alpha_{2}^{3/2}$, which would give a cubic equation in $z = \alpha_{2}^{3/2}$.


\subsection[Solutions for epsilon]{Solutions for $\epsilon_B$} \label{sec:app:scam:epsilonsol}

The function $\delta$ in \eq{eq:andeggcorr} depends on the parameter $e_B$ and involves its powers $e_{B}^{0}$, $e_{B}^{2}$, and $e_{B}^{4}$, but \eq{eq:freqeqapprox} itself contains $\epsilon_{B} = (1 - e_{B}^{2})^{-3/2}$.
Our goal is to achieve a simpler form of \eq{eq:freqeqapprox} that should only depend on $\epsilon_B$.
In fact, the MacLaurin series expansion of $\epsilon_B$ with respect to $e_B$ is
\[
    (1 - e_{B}^{2})^{-3/2} = 1 + \frac{3}{2} e_{B}^{2} + \frac{15}{4} e_{B}^{4} + \mathcal{O}(e_{B}^{6})
\]
and this expansion recovers the correct powers of $e_{B}$ in $\delta$.
There are other evidences that can justify the use of $\epsilon_{B}$ instead of $e_{B}$.
First, the correction $\delta$ extends the \citet{Hep1978} formula, which is a linear function of $\epsilon_{B}$.
Second, \citet{Geo2003} and \citet{Giu2011} constructed secular models for the three-body problem that extend Heppenheimer's model to higher order in the masses, and both new models contain some power of the expression $(1 - e_{B}^{2})$.
Following these arguments we set $\delta = \epsilon_{B} \left( \alpha_{2}^{3/2} \Gamma_{1} (\gamma_{2}) + \alpha_{2}^{9/2} \Gamma_{2} (\gamma_{2}) \right)$, where functions $\Gamma_1$ and $\Gamma_2$ collect all terms that depend on the parameter $\gamma_{2}$.
In this way $\delta$ becomes a linear function of $\epsilon_B$.
Then the full solutions are
\begin{equation} \label{eq:epsilonsol1}
    \epsilon_{B, \pm} =
    \frac{
        \gamma_{2} \alpha_{2}^{3} \Delta \pm
        \sqrt{
            (\gamma_{2} \alpha_{2}^{3} \Delta)^{2}
        - 4 (1 + \gamma_{1})^{1/2} \gamma_{1} \gamma_{2} \alpha_{1}^{3/2} \alpha_{2}^{9/2} \Gamma
        }
    }{2 (1 + \gamma_{1})^{1/2} \gamma_{2} \alpha_{2}^{9/2} \Gamma},
\end{equation}
with the abbreviations $\Delta = (1 + \gamma_{1})^{1/2} - \alpha_{1}^{3/2}$ ($\Delta > 0$) and $\Gamma = \Gamma_{1} + \alpha_{2}^{3} \Gamma_{2}$.
Note that $\epsilon_{B} \ge 1$ in order to be a physically relevant solution.


\section{Tables} \label{sec:app:tabs}


\begin{table}
    \centering
    \scriptsize
    \caption{G+F binary stars for different giant planet masses.}
    \label{tab:dataGF}

    \begin{tabular}{ r *6{c} }
\toprule

 & \multicolumn{2}{c}{$m_{GP} = 1$~M$_J$} &
   \multicolumn{2}{c}{$m_{GP} = 3$~M$_J$} &
   \multicolumn{2}{c}{$m_{GP} = 5$~M$_J$} \\

\cmidrule(lr){2-3} \cmidrule(lr){4-5} \cmidrule(lr){6-7}

 $a_{B}$ [AU] &
 $\alpha_{2,\text{IHZ}}$ & $\alpha_{2,\text{OHZ}}$ &
 $\alpha_{2,\text{IHZ}}$ & $\alpha_{2,\text{OHZ}}$ &
 $\alpha_{2,\text{IHZ}}$ & $\alpha_{2,\text{OHZ}}$ \\

\midrule
\multicolumn{7}{c}{$e_{B} = 0.0$} \\
\midrule

 20 & 0.088 & 0.120 & 0.103 & 0.135 & 0.112 & 0.144 \\
 50 & 0.058 & 0.074 & 0.071 & 0.088 & 0.078 & 0.096 \\
100 & 0.044 & 0.055 & 0.055 & 0.067 & 0.061 & 0.074 \\
200 & 0.034 & 0.042 & 0.043 & 0.053 & 0.048 & 0.058 \\
300 & 0.030 & 0.036 & 0.038 & 0.046 & 0.042 & 0.051 \\
400 & 0.027 & 0.033 & 0.034 & 0.042 & 0.038 & 0.046 \\
500 & 0.025 & 0.031 & 0.032 & 0.039 & 0.036 & 0.043 \\

\midrule
\multicolumn{7}{c}{$e_{B} = 0.2$} \\
\midrule

 20 & 0.087 & 0.119 & 0.102 & 0.134 & 0.110 & 0.141 \\
 50 & 0.057 & 0.073 & 0.070 & 0.087 & 0.077 & 0.095 \\
100 & 0.044 & 0.054 & 0.054 & 0.067 & 0.060 & 0.074 \\
200 & 0.034 & 0.042 & 0.043 & 0.052 & 0.048 & 0.058 \\
300 & 0.030 & 0.036 & 0.037 & 0.045 & 0.042 & 0.050 \\
400 & 0.027 & 0.033 & 0.034 & 0.041 & 0.038 & 0.046 \\
500 & 0.025 & 0.030 & 0.032 & 0.038 & 0.035 & 0.042 \\

\midrule
\multicolumn{7}{c}{$e_{B} = 0.4$} \\
\midrule

 20 & 0.083 & 0.114 & 0.100 & 0.126 & 0.104 & 0.132 \\
 50 & 0.055 & 0.070 & 0.067 & 0.083 & 0.073 & 0.090 \\
100 & 0.042 & 0.052 & 0.052 & 0.064 & 0.058 & 0.070 \\
200 & 0.033 & 0.040 & 0.041 & 0.050 & 0.046 & 0.055 \\
300 & 0.028 & 0.034 & 0.036 & 0.043 & 0.040 & 0.048 \\
400 & 0.026 & 0.031 & 0.032 & 0.039 & 0.036 & 0.044 \\
500 & 0.024 & 0.029 & 0.030 & 0.036 & 0.034 & 0.041 \\

\midrule
\multicolumn{7}{c}{$e_{B} = 0.6$} \\
\midrule

 20 & 0.076 & 0.104 & 0.100 & 0.113 & 0.100 & 0.118 \\
 50 & 0.050 & 0.064 & 0.060 & 0.074 & 0.065 & 0.080 \\
100 & 0.038 & 0.047 & 0.047 & 0.057 & 0.052 & 0.062 \\
200 & 0.029 & 0.036 & 0.037 & 0.045 & 0.041 & 0.049 \\
300 & 0.026 & 0.031 & 0.032 & 0.039 & 0.036 & 0.043 \\
400 & 0.023 & 0.028 & 0.029 & 0.035 & 0.033 & 0.039 \\
500 & 0.022 & 0.026 & 0.027 & 0.033 & 0.030 & 0.037 \\

\bottomrule
    \end{tabular}
\end{table}

\clearpage


\begin{table}
    \centering
    \scriptsize
    \caption{G+G binary stars for different giant planet masses}
    \label{tab:dataGG}

    \begin{tabular}{ r *6{c} }
\toprule

 & \multicolumn{2}{c}{$m_{GP} = 1$~M$_J$} &
   \multicolumn{2}{c}{$m_{GP} = 3$~M$_J$} &
   \multicolumn{2}{c}{$m_{GP} = 5$~M$_J$} \\

\cmidrule(lr){2-3} \cmidrule(lr){4-5} \cmidrule(lr){6-7}

 $a_{B}$ [AU] &
 $\alpha_{2,\text{IHZ}}$ & $\alpha_{2,\text{OHZ}}$ &
 $\alpha_{2,\text{IHZ}}$ & $\alpha_{2,\text{OHZ}}$ &
 $\alpha_{2,\text{IHZ}}$ & $\alpha_{2,\text{OHZ}}$ \\

\midrule
\multicolumn{7}{c}{$e_{B} = 0.0$} \\
\midrule

 20 & 0.092 & 0.125 & 0.108 & 0.141 & 0.118 & 0.150 \\
 50 & 0.061 & 0.077 & 0.075 & 0.093 & 0.082 & 0.102 \\
100 & 0.047 & 0.058 & 0.058 & 0.071 & 0.065 & 0.079 \\
200 & 0.036 & 0.044 & 0.046 & 0.056 & 0.051 & 0.062 \\
300 & 0.032 & 0.039 & 0.040 & 0.049 & 0.045 & 0.054 \\
400 & 0.029 & 0.035 & 0.036 & 0.044 & 0.041 & 0.049 \\
500 & 0.027 & 0.032 & 0.034 & 0.041 & 0.038 & 0.046 \\

\midrule
\multicolumn{7}{c}{$e_{B} = 0.2$} \\
\midrule

 20 & 0.091 & 0.124 & 0.107 & 0.139 & 0.116 & 0.148 \\
 50 & 0.060 & 0.077 & 0.074 & 0.092 & 0.081 & 0.101 \\
100 & 0.046 & 0.057 & 0.058 & 0.070 & 0.064 & 0.078 \\
200 & 0.036 & 0.044 & 0.045 & 0.055 & 0.051 & 0.061 \\
300 & 0.031 & 0.038 & 0.040 & 0.048 & 0.044 & 0.053 \\
400 & 0.028 & 0.034 & 0.036 & 0.044 & 0.040 & 0.049 \\
500 & 0.026 & 0.032 & 0.033 & 0.040 & 0.037 & 0.045 \\

\midrule
\multicolumn{7}{c}{$e_{B} = 0.4$} \\
\midrule

 20 & 0.087 & 0.118 & 0.102 & 0.131 & 0.110 & 0.139 \\
 50 & 0.058 & 0.074 & 0.070 & 0.088 & 0.077 & 0.095 \\
100 & 0.044 & 0.055 & 0.055 & 0.067 & 0.061 & 0.074 \\
200 & 0.034 & 0.042 & 0.043 & 0.053 & 0.048 & 0.058 \\
300 & 0.030 & 0.036 & 0.038 & 0.046 & 0.042 & 0.051 \\
400 & 0.027 & 0.033 & 0.034 & 0.042 & 0.038 & 0.046 \\
500 & 0.025 & 0.030 & 0.032 & 0.039 & 0.036 & 0.043 \\

\midrule
\multicolumn{7}{c}{$e_{B} = 0.6$} \\
\midrule

 20 & 0.079 & 0.108 & 0.100 & 0.118 & 0.100 & 0.123 \\
 50 & 0.052 & 0.067 & 0.063 & 0.078 & 0.069 & 0.085 \\
100 & 0.040 & 0.050 & 0.050 & 0.060 & 0.055 & 0.066 \\
200 & 0.031 & 0.038 & 0.039 & 0.047 & 0.044 & 0.052 \\
300 & 0.027 & 0.033 & 0.034 & 0.041 & 0.038 & 0.046 \\
400 & 0.025 & 0.030 & 0.031 & 0.038 & 0.035 & 0.042 \\
500 & 0.023 & 0.028 & 0.029 & 0.035 & 0.032 & 0.039 \\

\bottomrule
    \end{tabular}
\end{table}

\clearpage


\begin{table}
    \centering
    \scriptsize
    \caption{G+K binary stars for different giant planet masses}
    \label{tab:dataGK}

    \begin{tabular}{ r *6{c} }
\toprule

 & \multicolumn{2}{c}{$m_{GP} = 1$~M$_J$} &
   \multicolumn{2}{c}{$m_{GP} = 3$~M$_J$} &
   \multicolumn{2}{c}{$m_{GP} = 5$~M$_J$} \\

\cmidrule(lr){2-3} \cmidrule(lr){4-5} \cmidrule(lr){6-7}

 $a_{B}$ [AU] &
 $\alpha_{2,\text{IHZ}}$ & $\alpha_{2,\text{OHZ}}$ &
 $\alpha_{2,\text{IHZ}}$ & $\alpha_{2,\text{OHZ}}$ &
 $\alpha_{2,\text{IHZ}}$ & $\alpha_{2,\text{OHZ}}$ \\

\midrule
\multicolumn{7}{c}{$e_{B} = 0.0$} \\
\midrule

 20 & 0.098 & 0.132 & 0.116 & 0.150 & 0.126 & 0.161 \\
 50 & 0.065 & 0.083 & 0.081 & 0.100 & 0.089 & 0.110 \\
100 & 0.050 & 0.062 & 0.063 & 0.077 & 0.070 & 0.085 \\
200 & 0.039 & 0.048 & 0.050 & 0.060 & 0.056 & 0.067 \\
300 & 0.034 & 0.042 & 0.043 & 0.053 & 0.049 & 0.059 \\
400 & 0.031 & 0.038 & 0.039 & 0.048 & 0.044 & 0.053 \\
500 & 0.029 & 0.035 & 0.037 & 0.044 & 0.041 & 0.049 \\

\midrule
\multicolumn{7}{c}{$e_{B} = 0.2$} \\
\midrule

 20 & 0.097 & 0.130 & 0.115 & 0.148 & 0.125 & 0.158 \\
 50 & 0.065 & 0.082 & 0.080 & 0.099 & 0.088 & 0.108 \\
100 & 0.050 & 0.061 & 0.062 & 0.076 & 0.069 & 0.084 \\
200 & 0.039 & 0.047 & 0.049 & 0.060 & 0.055 & 0.066 \\
300 & 0.034 & 0.041 & 0.043 & 0.052 & 0.048 & 0.058 \\
400 & 0.031 & 0.037 & 0.039 & 0.047 & 0.044 & 0.053 \\
500 & 0.028 & 0.035 & 0.036 & 0.044 & 0.040 & 0.049 \\

\midrule
\multicolumn{7}{c}{$e_{B} = 0.4$} \\
\midrule

 20 & 0.093 & 0.124 & 0.109 & 0.140 & 0.118 & 0.148 \\
 50 & 0.062 & 0.079 & 0.076 & 0.094 & 0.084 & 0.103 \\
100 & 0.048 & 0.059 & 0.060 & 0.073 & 0.066 & 0.080 \\
200 & 0.037 & 0.045 & 0.047 & 0.057 & 0.052 & 0.063 \\
300 & 0.032 & 0.039 & 0.041 & 0.050 & 0.046 & 0.055 \\
400 & 0.029 & 0.036 & 0.037 & 0.045 & 0.042 & 0.050 \\
500 & 0.027 & 0.033 & 0.035 & 0.042 & 0.039 & 0.047 \\

\midrule
\multicolumn{7}{c}{$e_{B} = 0.6$} \\
\midrule

 20 & 0.084 & 0.113 & 0.100 & 0.125 & 0.105 & 0.131 \\
 50 & 0.056 & 0.071 & 0.068 & 0.084 & 0.075 & 0.091 \\
100 & 0.043 & 0.053 & 0.054 & 0.065 & 0.059 & 0.072 \\
200 & 0.034 & 0.041 & 0.043 & 0.051 & 0.047 & 0.057 \\
300 & 0.029 & 0.036 & 0.037 & 0.045 & 0.041 & 0.050 \\
400 & 0.027 & 0.032 & 0.034 & 0.041 & 0.038 & 0.045 \\
500 & 0.025 & 0.030 & 0.031 & 0.038 & 0.035 & 0.042 \\

\bottomrule
    \end{tabular}
\end{table}

\clearpage


\begin{table}
    \centering
    \scriptsize
    \caption{G+M binary stars for different giant planet masses}
    \label{tab:dataGM}

    \begin{tabular}{ r *6{c} }
\toprule

 & \multicolumn{2}{c}{$m_{GP} = 1$~M$_J$} &
   \multicolumn{2}{c}{$m_{GP} = 3$~M$_J$} &
   \multicolumn{2}{c}{$m_{GP} = 5$~M$_J$} \\

\cmidrule(lr){2-3} \cmidrule(lr){4-5} \cmidrule(lr){6-7}

 $a_{B}$ [AU] &
 $\alpha_{2,\text{IHZ}}$ & $\alpha_{2,\text{OHZ}}$ &
 $\alpha_{2,\text{IHZ}}$ & $\alpha_{2,\text{OHZ}}$ &
 $\alpha_{2,\text{IHZ}}$ & $\alpha_{2,\text{OHZ}}$ \\

\midrule
\multicolumn{7}{c}{$e_{B} = 0.0$} \\
\midrule

 20 & 0.108 & 0.144 & 0.130 & 0.166 & 0.142 & 0.179 \\
 50 & 0.073 & 0.092 & 0.091 & 0.112 & 0.101 & 0.124 \\
100 & 0.057 & 0.070 & 0.071 & 0.087 & 0.079 & 0.097 \\
200 & 0.045 & 0.054 & 0.056 & 0.068 & 0.063 & 0.076 \\
300 & 0.039 & 0.047 & 0.049 & 0.060 & 0.055 & 0.066 \\
400 & 0.035 & 0.043 & 0.045 & 0.054 & 0.050 & 0.060 \\
500 & 0.033 & 0.040 & 0.042 & 0.050 & 0.046 & 0.056 \\

\midrule
\multicolumn{7}{c}{$e_{B} = 0.2$} \\
\midrule

 20 & 0.107 & 0.142 & 0.128 & 0.164 & 0.140 & 0.176 \\
 50 & 0.072 & 0.091 & 0.090 & 0.111 & 0.100 & 0.122 \\
100 & 0.056 & 0.069 & 0.071 & 0.086 & 0.078 & 0.095 \\
200 & 0.044 & 0.054 & 0.056 & 0.067 & 0.062 & 0.075 \\
300 & 0.038 & 0.047 & 0.049 & 0.059 & 0.054 & 0.066 \\
400 & 0.035 & 0.042 & 0.044 & 0.053 & 0.049 & 0.060 \\
500 & 0.032 & 0.039 & 0.041 & 0.050 & 0.046 & 0.055 \\

\midrule
\multicolumn{7}{c}{$e_{B} = 0.4$} \\
\midrule

 20 & 0.103 & 0.136 & 0.122 & 0.154 & 0.132 & 0.165 \\
 50 & 0.069 & 0.087 & 0.086 & 0.106 & 0.095 & 0.116 \\
100 & 0.054 & 0.066 & 0.067 & 0.082 & 0.075 & 0.091 \\
200 & 0.042 & 0.051 & 0.053 & 0.064 & 0.059 & 0.072 \\
300 & 0.037 & 0.044 & 0.046 & 0.056 & 0.052 & 0.063 \\
400 & 0.033 & 0.040 & 0.042 & 0.051 & 0.047 & 0.057 \\
500 & 0.031 & 0.037 & 0.039 & 0.047 & 0.044 & 0.053 \\

\midrule
\multicolumn{7}{c}{$e_{B} = 0.6$} \\
\midrule

 20 & 0.093 & 0.124 & 0.109 & 0.138 & 0.118 & 0.146 \\
 50 & 0.063 & 0.080 & 0.077 & 0.095 & 0.085 & 0.103 \\
100 & 0.049 & 0.060 & 0.061 & 0.074 & 0.068 & 0.082 \\
200 & 0.038 & 0.047 & 0.048 & 0.058 & 0.054 & 0.065 \\
300 & 0.033 & 0.040 & 0.042 & 0.051 & 0.047 & 0.057 \\
400 & 0.030 & 0.037 & 0.038 & 0.046 & 0.043 & 0.052 \\
500 & 0.028 & 0.034 & 0.036 & 0.043 & 0.040 & 0.048 \\

\bottomrule
    \end{tabular}
\end{table}

\clearpage







\bibliographystyle{apalike}
\bibliography{references} 






\end{document}